\documentclass[twocolumn]{aastex62}

\usepackage{natbib}
\usepackage{multirow}
\usepackage{listings}
\usepackage{color}
\usepackage{xcolor}

\definecolor{dkgreen}{rgb}{0,0.6,0}
\definecolor{gray}{rgb}{0.5,0.5,0.5}
\definecolor{mauve}{rgb}{0.58,0,0.82}

\submitjournal{ApJ}

\shorttitle{DASH}
\shortauthors{Muthukrishna et al.}

\begin{document}

\title{DASH: Deep Learning for the Automated Spectral Classification of Supernovae and their Hosts}

\correspondingauthor{Daniel Muthukrishna}
\email{daniel.muthukrishna@ast.cam.ac.uk}

\author[0000-0002-5788-9280]{Daniel Muthukrishna}
\affiliation{Institute of Astronomy, University of Cambridge, Madingley
Road, Cambridge CB3 0HA, UK}
\affiliation{The Research School of Astronomy and Astrophysics, Australian National University, ACT 2601, Australia}
\affiliation{ARC Centre of Excellence for All-sky Astrophysics (CAASTRO)}

\author[0000-0002-7464-2351]{David Parkinson}
\affiliation{ARC Centre of Excellence for All-sky Astrophysics (CAASTRO)}
\affiliation{School of Mathematics and Physics, University of Queensland, Brisbane, QLD 4072, Australia}
\affiliation{Korea Astronomy and Space Science Institute, 776, Daedeokdae-ro, Yuseong-gu, Daejeon 34055, Republic of Korea}
\author[0000-0002-4283-5159]{Brad E. Tucker}
\affiliation{Mt Stromlo Observatory, The Research School of Astronomy and Astrophysics, Australian National University, ACT 2601, Australia}
\affiliation{National Centre for the Public Awareness of Science, the Australian National University, Canberra, Australia}
\affiliation{ARC Centre of Excellence for All-sky Astrophysics (CAASTRO)}

\begin{abstract}
We present {\tt DASH} (Deep Automated Supernova and Host classifier), a novel software package that automates the classification of the type, age, redshift, and host galaxy of supernova spectra. {\tt DASH} makes use of a new approach that does not rely on iterative template matching techniques like all previous software, but instead classifies based on the learned features of each supernova's type and age. It has achieved this by employing a deep convolutional neural network to train a matching algorithm. This approach has enabled {\tt DASH} to be orders of magnitude faster than previous tools, being able to accurately classify hundreds or thousands of objects within seconds. We have tested its performance on four years of data from the Australian Dark Energy Survey (OzDES). The deep learning models were developed using {\tt TensorFlow}, and were trained using over 4000 supernova spectra taken from the CfA Supernova Program and the Berkeley SN Ia Program as used in {\tt SNID} (Supernova Identification software, \citealt{Blondin2007}). \textcolor{black}{Unlike template matching methods, }the trained models are independent of the number of \textcolor{black}{spectra in the training data}, which allows for {\tt DASH}'s unprecedented speed. We have developed both a graphical interface for easy visual classification and analysis of supernovae, and a {\tt Python} library for the autonomous and quick classification of several supernova spectra. The speed, accuracy, user-friendliness, and versatility of {\tt DASH} presents an advancement to existing spectral classification tools. We have made the code publicly available on {\tt GitHub} and PyPI ({\tt pip install astrodash}) to allow for further contributions and development. The package documentation is available at \url{https://astrodash.readthedocs.io}.

\end{abstract}

\keywords{methods: data analysis, machine learning, statistical --- supernovae: general --- surveys --- techniques: spectroscopic --- cosmology: observations}

\section{Introduction}
Supernovae (SNe) have been pivotal to modern observational cosmology. The use of Type Ia supernovae (SNIa) as standard candles have provided some of the most compelling evidence for the discovery that the expansion of the universe is accelerating \citep{Riess1998,Perlmutter1999,Schmidt1998}. However, the nature of dark energy and the value of many cosmological parameters are still under active consideration \citep{Zhang2017,Muthukrishna2016}. To this end, several large scale surveys including the Dark Energy Survey (DES) \citep{DarkEnergySurveyCollaboration2016}, the Supernova Legacy Survey (SNLS) \citep{Astier2006}, and ESSENCE \citep{Davis2007} have aimed to increase the total set of supernovae in order to gain a better understanding of dark energy. Moreover, in the near future, projects such as the Large Synoptic Survey Telescope \citep[LSST, ][]{LSSTScienceCollaborations2009} will substantially increase the transient catalogue with the expectation to observe orders of magnitude more supernovae than ever before. 

The field of observational astronomy has reached a new era of `big data', where we are collecting more data than humans can possibly process and classify alone. Machine learning techniques have been a key driver in tackling these new large-scale problems and many successful attempts have been used to solve large data astronomy problems \citep{Ball2009}. More recently, however, deep learning has gained a lot of popularity in the machine learning community for its accuracy, efficiency and flexibility. In particular, convolutional neural networks (CNNs) have achieved remarkable results in a range of different applications including image and speech recognition challenges, outperforming previous approaches (e.g. \citet{Krizhevsky2012,Razavian2014,Szegedy2015}). Only after the Galaxy Zoo Challenge \citep{Lintott2008,Dieleman2015}, however, did it begin to gain a larger interest in the astronomy community (e.g. \citet{Cabrera-Vives2017,Aniyan2017}). 

While machine learning has been applied to photometric supernova classification \citep[e.g. ][in prep.]{Moller2016,Lochner2016,Charnock2016,Moss2018,Narayan2018,Muthukrishna18}, few attempts at spectral classification of any kind have been made. While this project was being developed, a paper by \citet{Sasdelli2015} applied deep learning to supernova spectra: using it to explore the spectroscopic diversity in Type-Ia supernovae. Moreover, a recent thesis by \citet{Hala2014} has applied a similar CNN approach to that described in this paper to the spectral classification of quasars, stars and galaxies. Supernovae are inherently more complicated, however, due to the fact that they vary with time and have degeneracies in their type, age and redshift, with often lower signal-to-noise caused by distortions from their host galaxy.

In fact, there are several factors that make supernova classification a challenging problem. While different types of supernovae are distinguished by the presence of particular absorption features in their spectra, the problem of spectral classification is made difficult by the fact that the spectrum changes depending on the number of days since maximum light it was observed at (defined as `age' in this paper). Each spectrum also has distortions due to contamination from host galaxy light. Moreover, the redshift at which the supernova is observed impacts which spectral features are visible in the observed wavelength range, and also affects the signal-to-noise ratio (S/N) which decreases with redshift. Extinction from interstellar dust further impacts the spectra. Subtracting the continuum from each spectrum can limit this issue by placing more emphasis on the spectral features instead of the colour information. Finally, issues with the telescope used to observe the spectrum such as dichroic jumps being caused by miscalibrations between the two spectral arms using different CCDs, and also telluric features from the earth's atmosphere are further problems that need to be accounted for when classifying spectra.

\subsection{Prior Software}

Due to these complications, existing supernova spectral classifiers are not able to automate the classification process. Currently, the process of classifying supernovae is very slow and labour-intensive, with the classification process for a single supernova taking up to a few hours with the incessant input of an experienced astronomer. Surveys like the Australian Dark Energy Survey \citep[OzDES, ][]{Yuan2015,Childress2017} are observing thousands of transient objects which need to be classified; and current methods make this an enormously time-consuming process. {\tt SNID} \citep{Blondin2007} and {\tt Superfit} \citep{Howell2005} are the two main spectral classifier software packages used to classify supernovae. {\tt SNID} is a fast typing tool written in {\tt Fortran}. It makes use of the cross-correlation algorithms of \citet{Tonry1979} and has been effective in distinguishing SN subtypes at a range of redshifts. However, it's accuracy drops significantly when there is host-galaxy contamination or if the spectra has a low S/N. In such cases, {\tt Superfit} acts as a better tool due to its ability to classify host-contaminated spectra, and account for extinction, and as such is the primary tool used by large surveys such as OzDES and SNLS. It's main downfall, however, is that it is often very slow and requires a lot of user-input to constrain priors on redshift, host, and supernova type. {\tt Superfit} is written in {\tt IDL} and makes use of a chi-squared minimisation approach to classify the spectra. It accounts for the supernova type, age, host galaxy and extinction in its minimisation equation, which enable it to be a very effective tool. However, given the thousands of transient objects that are being detected by the latest era of supernova surveys, a faster and more autonomous software is required. 

{\tt DASH} makes use of the techniques used in each of these previous tools. In particular, the spectra in {\tt DASH} are processed in a very similar method to the log-wavelength spectra developed by {\tt SNID} (see section \ref{sec:Preprocessing}). Moreover, the $rlap$ ranking system developed by \citet{Blondin2007} is available in {\tt DASH} and is used as a test for misclassifications (along with the machine learning scores) in much the same way as {\tt SNID}. 

All previous spectral tools for classification and redshifting make use of the \citet{Tonry1979} cross-correlation technique (i.e. {\tt SNID}, {\tt MARZ} \citep{Hinton}, {\tt AUTOZ} \citep{Baldry2014}, {\tt RUNZ}) or a chi-squared minimisation approach (i.e. {\tt Superfit}). However, using either of these techniques means that the total computation time increases linearly with the number of \textcolor{black}{spectra in the dataset}. Both {\tt SNID} and {\tt Superfit} can only compare an input spectrum with one \textcolor{black}{other spectrum} at a time, and their accuracy is highly reliant on their dataset. {\tt DASH} improves upon this by using the aggregate features of a particular class of supernova instead of comparing to a single spectrum. {\tt DASH} is able to learn from the features of all spectra in a supernova class and classify on that, instead of comparing to just one spectrum at a time like previous tools.


\subsection{Overview}
We have developed a new supernova spectral classification tool, {\tt DASH} (Deep Automated Supernova and Host classifier), to quickly and accurately determine the type, age, redshift, and host galaxy of supernova spectra. We make use of a convolutional neural network which greatly improves upon many aspects of previous classification tools. \textcolor{black}{In section \ref{sec:Data}, we detail} the datasets we have collated and the pre-processing techniques that are uniformly applied to the spectra. In section \ref{sec:DeepLearning}, we describe the convolutional neural network architecture that we use. In section \ref{sec:TrainedModels}, we outline the four different trained models that are available in the {\tt DASH} release, before describing the algorithms used to redshift and to warn the user against possible misclassifications. In Appendix \ref{sec:IntendedUse}, we outline how to use the {\tt Python} library and graphical interfaces, as well as detail the platform requirements and the code development. Finally, in section \ref{sec:Performance}, we evaluate the performance of {\tt DASH} on a validation set and the recent OzDES data.

\section{Data}
\label{sec:Data}
Supernovae are the result of either the core-collapse of massive stars or the thermonuclear disruption of carbon-oxygen white dwarfs accreting matter from a binary companion. They are classified based on the presence of certain features in their optical spectrum taken near maximum light instead of their explosion mechanism. 
The presence or absence of Hydrogen, Silicon and Helium spectral features separate SNe into four broad types: Type-Ia (SNIa), Type-Ib (SNIb), Type-Ic (SNIc), and Type-II (SNII). Within each of these, several subtypes have been defined due to a range of peculiarities in their spectra. {\tt DASH} makes use of 17 subtypes defined by \citet{Blondin2007}, \citet{Modjaz2016}, and \citet{Silverman2012}:
\begin{description}
	\item [SNIa] Ia-norm, Ia-pec, Ia-91T, Ia-91bg, Ia-csm, Iax
	\item [SNIb] Ib-norm, Ib-pec, Ib-n, IIb
	\item [SNIc] Ic-norm, Ic-pec, Ic-broad
	\item [SNII] IIP, II-pec, IIL, IIn
\end{description}

In order to train the model, it was important that we collected a wide range of spectra encompassing each of these subtypes over a range of different ages. The quality of the classification model is highly dependent on the data that it was trained on, and hence, in this section we detail how the data was collected, outline the decisions made that led to the final dataset, and describe the systematic pre-processing techniques applied to the data before it was trained using a deep convolutional neural network (CNN).

\subsection{Description}
\label{sec:Data_Description}
We collected labelled spectra from three main repositories: the {\tt SNID} database, the Berkeley Supernovae Ia Program (BSNIP), and the releases from Liu \& Modjaz in 2014-2016.

\subsubsection{{\tt SNID} Database}
The latest version of the {\tt SNID} database (Templates 2.0\footnote{\url{https://people.lam.fr/blondin.stephane/software/snid/index.html}}) has compiled 3716 spectra from 333 different supernovae obtained from 1979 to 2008 \citep{Blondin2007,Blondin2012}. These were collected from the SUSPECT public archive\footnote{\url{http://bruford.nhn.ou.edu/~suspect/index1.html}}, the CfA Supernova Archive\footnote{\url{http://www.cfa.harvard.edu/supernova/SNarchive.html}}, and the CfA Supernova Program \citep{Matheson2008,Blondin2012}. The collected set were selected to have a high signal-to-noise ratio and have been cleaned, de-redshifted, continuum-divided, smoothed and processed onto a log-wavelength scale by {\tt SNID} in a process defined by \citet{Blondin2007}. The spectra were classed into 14 different subtypes: Ia-norm, Ia-pec, Ia-91T, Ia-91bg, Ia-csm, Ib-norm, Ib-pec, IIb, Ic-norm, Ic-broad, IIP, II-pec, IIL, IIn. A detailed description of these subtypes can be found in \citet{Blondin2007}.

We removed the supernovae where the date of maximum light was unknown, and were left with a total of 3618 spectra from 317 different SNe. This distribution comprised of 2724 spectra from 283 SNIa, 223 spectra from 12 SNIb, 183 spectra from 11 SNIc, and 488 spectra from 11 SNII.

\subsubsection{Liu \& Modjaz}
In 2014-2016, Yuqian Liu and Maryam Modjaz released a series of papers \citep{Modjaz2016,Modjaz2014a,Liu,Liu2015} which collected the largest set of stripped-envelope core-core collapse supernovae (SNIb and SNIc). The spectral database was downloaded from their {\tt GitHub} repository\footnote{\url{https://github.com/nyusngroup/SESNtemple/tree/master/SNIDtemplates}} and contained 1045 spectra across 96 Type Ib and Ic SNe. Within this set, \citet{Liu2015} corrected 14 SNe also included in the {\tt SNID} Templates 2.0 release, which had incorrect type or age information. In addition, they introduced two new subtypes called Ib-n \citep[defined in ][]{Pastorello2008} and Ic-pec to better account for variations in some spectra. 

We again removed the supernovae where the date of maximum light was unknown, and were left with a total of  571 spectra from 57 SNe. The distribution comprised of 323 spectra from 27 SNIb, 248 spectra from 30 SNIc, and zero SNIa or SNII spectra.

\subsubsection{BSNIP}
In 2012, \citet{Silverman2012} collated 1126 spectra from 277 supernovae as part of the Berkeley SN Ia Program (BSNIP) in the BSNIP v7.0 release\footnote{\url{https://people.lam.fr/blondin.stephane/software/snid/index.html}}. Many of these were, however, also part of the {\tt SNID} Templates 2.0 set and the Liu \& Modjaz updates. After removing exact duplicates in the BSNIP v7.0 database and also removing spectra with an unknown date of maximum light, we were left with 604 spectra across 133 SNe. This reduced set had 29 new SNe and 114 SNe that were common to the previously discussed datasets but included spectra at different ages. The distribution comprised of 564 spectra from 131 SNIa, 40 spectra from 2 SNIc, and zero SNIb or SNII spectra.

The BSNIP release also defined two new subtypes called Ia-02cx (renamed Iax) and Ia-99aa \citep[defined in ][]{Silverman2012,Foley2013-Iax}. We discussed these subtypes with the author, Jeffrey Silverman, and he believed that Ia-99aa's are a subset of the Ia-91T type, and may not need their own category. Based on this discussion, and the fact that there were not enough Ia-99aa spectra to train its own subtype, we reclassified the Ia-99aa spectra as Ia-91T SNe.

While duplicate spectra between the datasets were removed, wherever there were discrepancies in the phase or subtype of a SN, we preferentially selected the Liu \& Modjaz spectra as this dataset was released the latest and also purposefully corrected the {\tt SNID} Templates 2.0 release. There were a total of six discrepancies in the subtypes of SNe from the BSNIP v7.0 and the {\tt SNID} Templates 2.0 datasets. The subtypes from the BSNIP dataset were selected in favour of the Templates 2.0 dataset because BSNIP intentionally improved upon the {\tt SNID} dataset. The following changes were made: sn2002cx, sn2005hk, sn2008A from Templates 2.0 were changed from Ia-pec to Iax subtypes; and the sn1995ac, sn2000cn, and sn2004aw from Templates 2.0 were changed to Ia-91T, Ia-91bg, and Ic-pec from the norm subtypes, respectively.

\subsection{{\tt DASH} data distribution}
\label{sec:template_distribution}
Combining the spectra from the {\tt SNID} Templates 2.0 database, the Liu \& Modjaz updates, and the BSNIP v7.0 release, and removing spectra with unknown ages, we were left with a  total of 4831 unique spectra across 403 unique SNe. The distribution comprised of 3288 spectra from 312 SNIa, 550 spectra from 40 SNIb, 505 spectra from 40 SNIc, and 488 spectra from 11 SNII.

\begin{figure}[ht!]
	\includegraphics[width=1\linewidth]{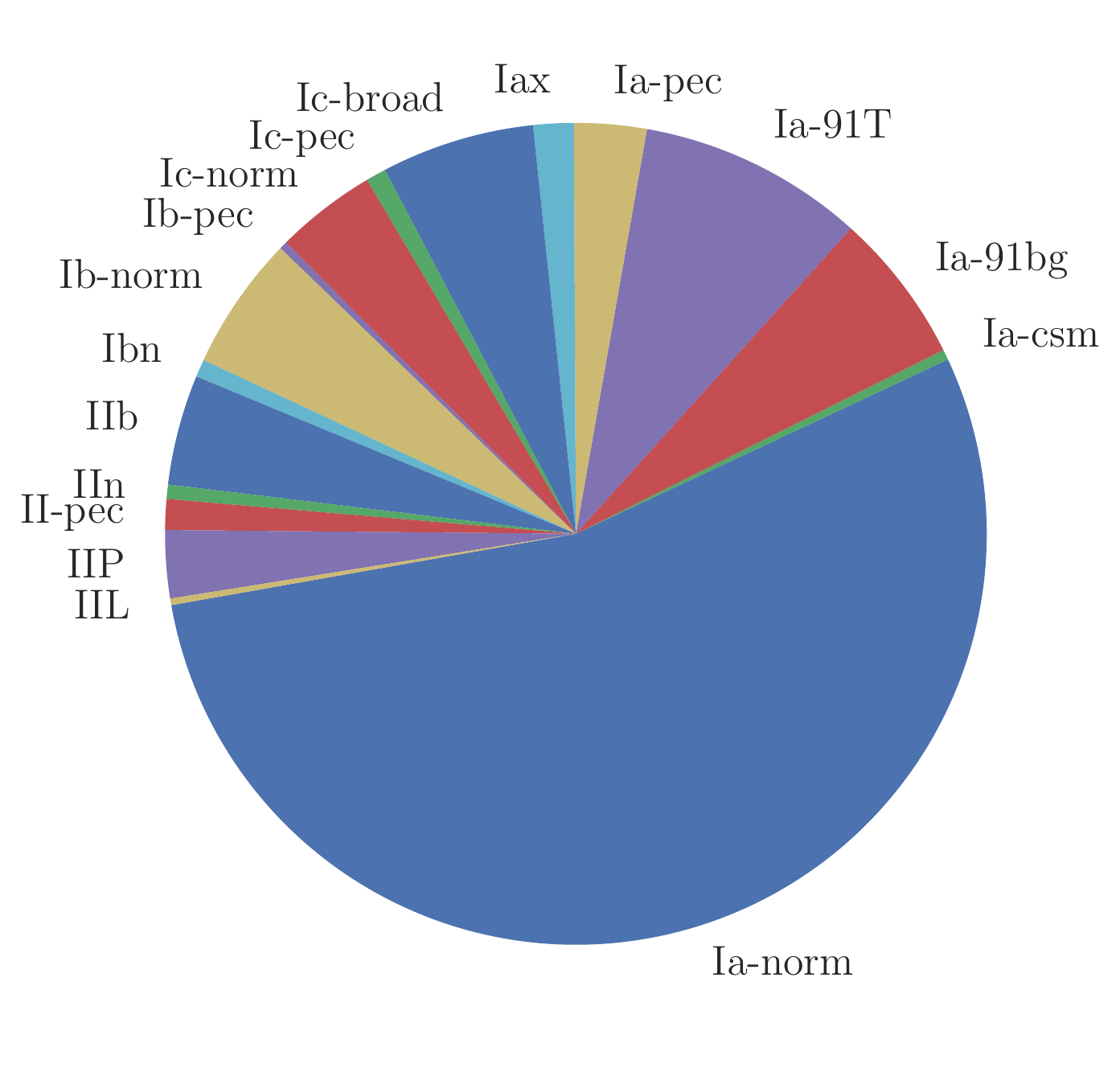}
	\caption{The fraction of spectra for each subtype in the final dataset. The total distribution separated by subtype and age is listed in Appendix \ref{sec:Datadistribution}, Figure \ref{fig:templatedistribution}.}
	\label{fig:type_counts_pie_chart}
\end{figure}

\begin{figure}[ht!]
	\includegraphics[width=1\linewidth]{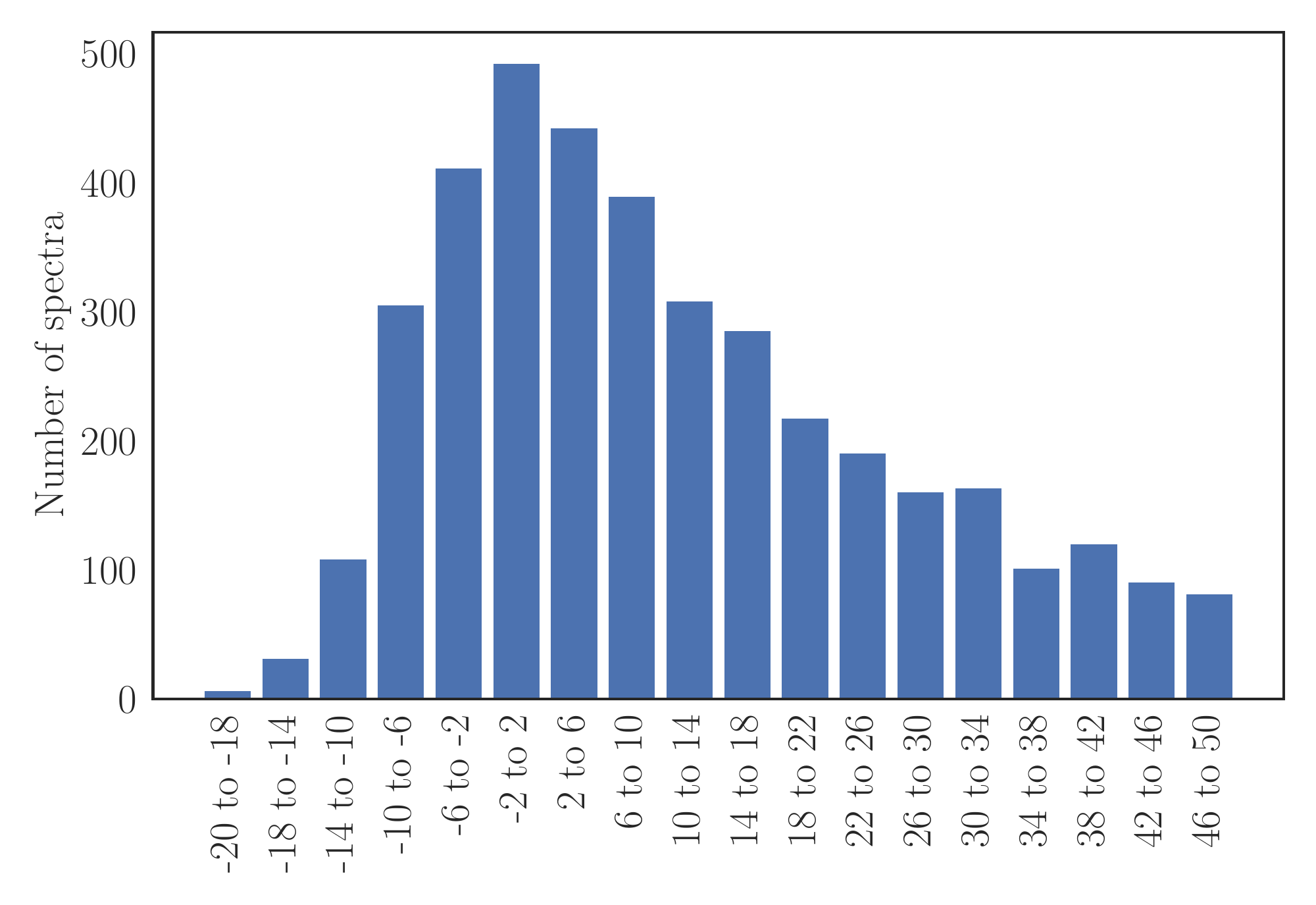}
	\caption{The distribution of spectra in the final dataset across the different age bins.}
	\label{fig:ages_counts_bar_chart}
\end{figure}

In general, supernovae that are observed several weeks before or after maximum light are usually very dim, and their spectra are mostly dominated by host galaxy light. Thus, we only considered supernovae between the range of -20 days to +50 days since maximum light. After removing spectra outside this range, we were left with 3899 spectra from 403 SNe. In order to group the spectra into bins that can be trained on for the machine learning algorithm, we split the ages into 4-day intervals. Therefore, for each of the 17 supernova subtypes, there are 18 age bins, leading to a total of 306 different classes to separate all of the spectra. \textcolor{black}{The distribution of spectra across the supernova subtypes and ages are illustrated in Figures \ref{fig:type_counts_pie_chart} and \ref{fig:ages_counts_bar_chart}, respectively. The complete distribution in each type and age classification bin is listed in Appendix \ref{sec:Datadistribution}, Figure \ref{fig:templatedistribution}}.

\subsubsection{Data Augmentation}
\label{sec:AugmentationAndOversampling}
Figures  \ref{fig:type_counts_pie_chart}, \ref{fig:ages_counts_bar_chart}, and \ref{fig:templatedistribution} illustrate two significant problems. Firstly, there are several bins with zero spectra, meaning that {\tt DASH} (and previous tools such as {\tt Superfit} and {\tt SNID}) will never be able to classify a spectrum into this bin. There is no way to fix this problem other than to observe a wider range of supernovae. In fact, while the dataset has proven to be sufficient for effective classification (see section \ref{sec:Performance}), it is expected that if a wider and deeper range of spectra is added to the training set, the accuracy - particularly for low-S/N spectra - will improve. 

Secondly, the rarity of some SN types, and the bias of cosmological surveys to preferentially observe Type-Ia supernovae near maximum-light over other types mean that there is a large imbalance in the dataset. In {\tt SNID} and {\tt Superfit}, this leads to a `type attractor' \citep{Blondin2007}, whereby low-S/N spectra will preferentially be classified as SNIa regardless of their actual type, simply because there are more SNIa spectra to choose from. Moreover, while this is a large set of supernovae, in terms of standard machine learning problems, this is a relatively small dataset. In order to combat both of these issues, we have made use of an oversampling technique to greatly diminish the effect of these problems. The idea of oversampling is to repeat each spectrum in lowly populated bins until all classification bins have the same number of spectra. As an example, if a bin in the training set had 250 spectra in it, we would repeat each of those spectra 4 times; and if a bin had 5 spectra, we repeat each of those spectra 200 times; until all bins have an equal amount of 1000 spectra. However, instead of simply repeating spectra (which adds no information to a neural network), we perform three data augmentation techniques which can magnify the size of our training set by over 1000 times. The following data augmentation steps are used:
\begin{description}
	\item[Adding noise] \hfill \\ 
	    The easiest thing to do is to simply add random amounts of Gaussian noise to each spectrum while oversampling. \textcolor{black}{In our case, we add Gaussian noise $\mathcal{N}(\mu, \sigma^2)$ with mean $\mu = 0$ and sigma $\sigma=0.05(f_{max} - f_{min})$, where $f_{max}$ and  $f_{min}$ are the maximum and minimum flux values in the spectrum, respectively.}
	\item[Adding host galaxy spectra] \hfill \\ 
    	Second, so that we can also distinguish a supernova spectrum which is contaminated by its host galaxy, we also add on varying amounts of host galaxy spectra. For each spectrum in the initial training set, we add a host galaxy spectrum in varying proportions from 1\% to 99\%, and also make use of 11 different host types: E, S0, Sa, Sb, Sc, SB1, SB2, SB3, SB4, SB5, SB6 which are taken from the BSNIP and {\tt Superfit} \textcolor{black}{datasets}.
	\item[Cropping] \hfill \\ 
    	We also crop each spectrum by varying amounts, such that instead of just training on an entire spectrum, we train on different wavelength segments of each spectrum. \textcolor{black}{That is, we reduce the wavelength range of each spectrum by random amounts. As with all spectra that do not cover the full wavelength range used in our neural network, we set the points in the preprocessed and normalised spectra that do not have data to 0.5 (see Figure \ref{fig:preprocessing}d for an example and section \ref{sec:Preprocessing} for more details).}

	\item[Redshifting] \hfill \\ 
	      Finally, for the unknown redshift models (see section \ref{sec:Models}) we redshift each spectrum by random amount from $z=0$ to $z=1$.
\end{description}

These processes increase the size of our training set considerably. Since we add on 11 different host galaxy spectra at over 10 different fractions, crop each spectrum at at least 4 different wavelength intervals, and add noise to each spectrum while oversampling by a minimum of 4 times (up to 1000 times depending on the number of spectra in the bin), we effectively increase our training set by at least $11 \times 10 \times 10 \times 4 \times 4 = 1760$ times the initial training set, but actually over around 100000 times the initial dataset size, given the amount of oversampling of lowly populated bins, and random redshifting during training.

In this data augmentation process, we are enabling the neural network to find and train on the common features among the augmented spectra, allowing it to train only on the actual features that make up a spectrum instead of the noise, host light, or wavelength range of each spectrum. This significantly inhibits the imbalanced dataset problem, and allows the neural network to train on actual SN features of a particular classification bin rather than random distortions of a single spectrum.

Ultimately, this technique is very important and effective, but can't compete with actually having huge amounts of real observational data. In future, as more large scale surveys work to increase the transient catalogue, these CNN problems will be far more powerful than what can be made with current datasets. 

\textcolor{black}{Before augmentation, we split the total set of transients into
two parts: 80\% for the \textit{training set} and 40\% for the \textit{testing set}. The \textit{training set} is used to train the classifier to
identify the correct supernovae class, while the \textit{testing set}
is used to test the performance of the classifier. We then apply the augmentation detailed previously to the training set only.}


\subsection{Preprocessing}
\label{sec:Preprocessing}
Arguably one of the most important aspects in an effective learning algorithm is the quality of the training set. As such, a lot of the software effort in this project has been in ensuring that the data has been processed in a systematic and uniform way before we train the matching algorithm. In this section, we outline the processing techniques used to prepare the training set and the input spectra.

\begin{figure*}
	\gridline{
		\fig{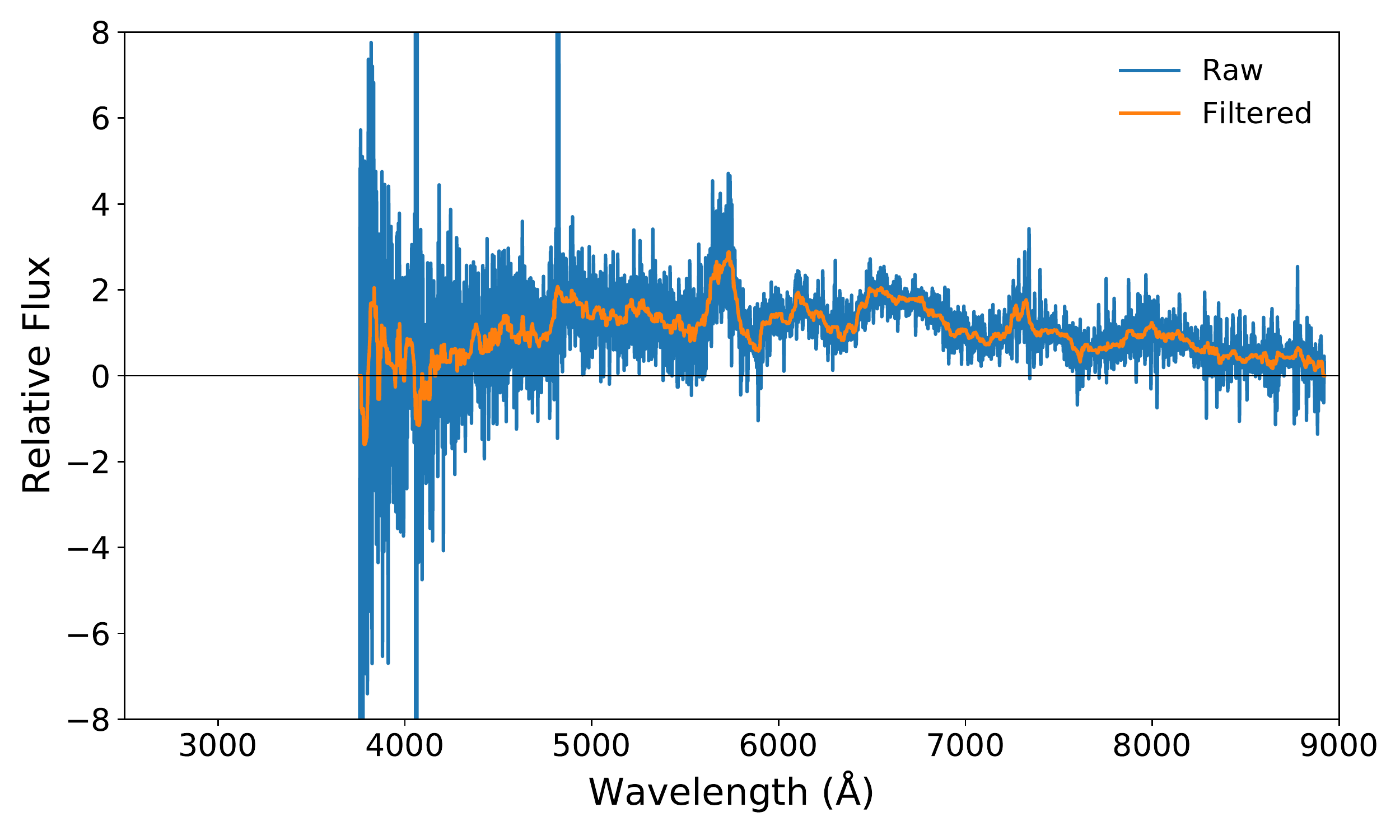}{0.5\textwidth}{(a)}
		\fig{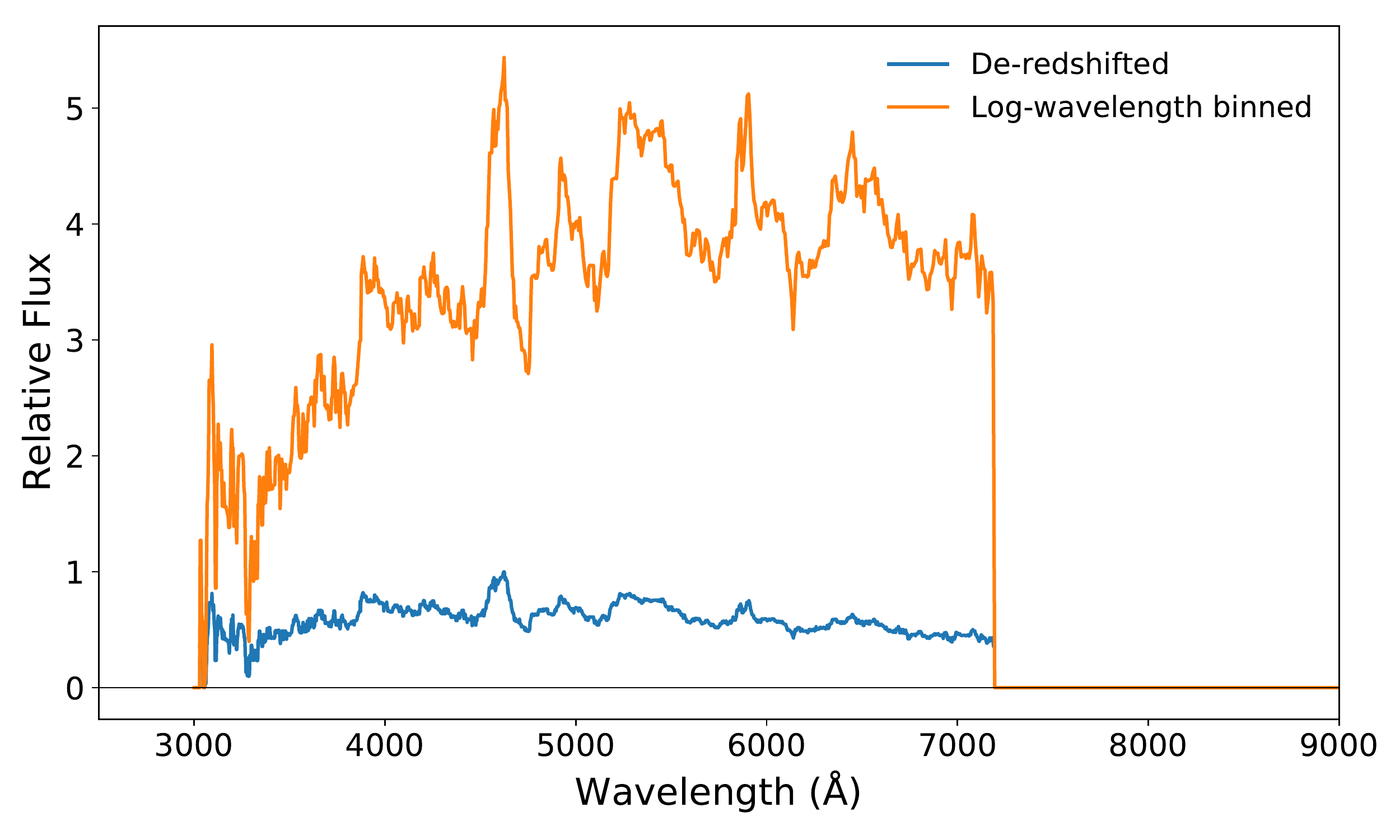}{0.5\textwidth}{(b)}
	}
	\gridline{
		\fig{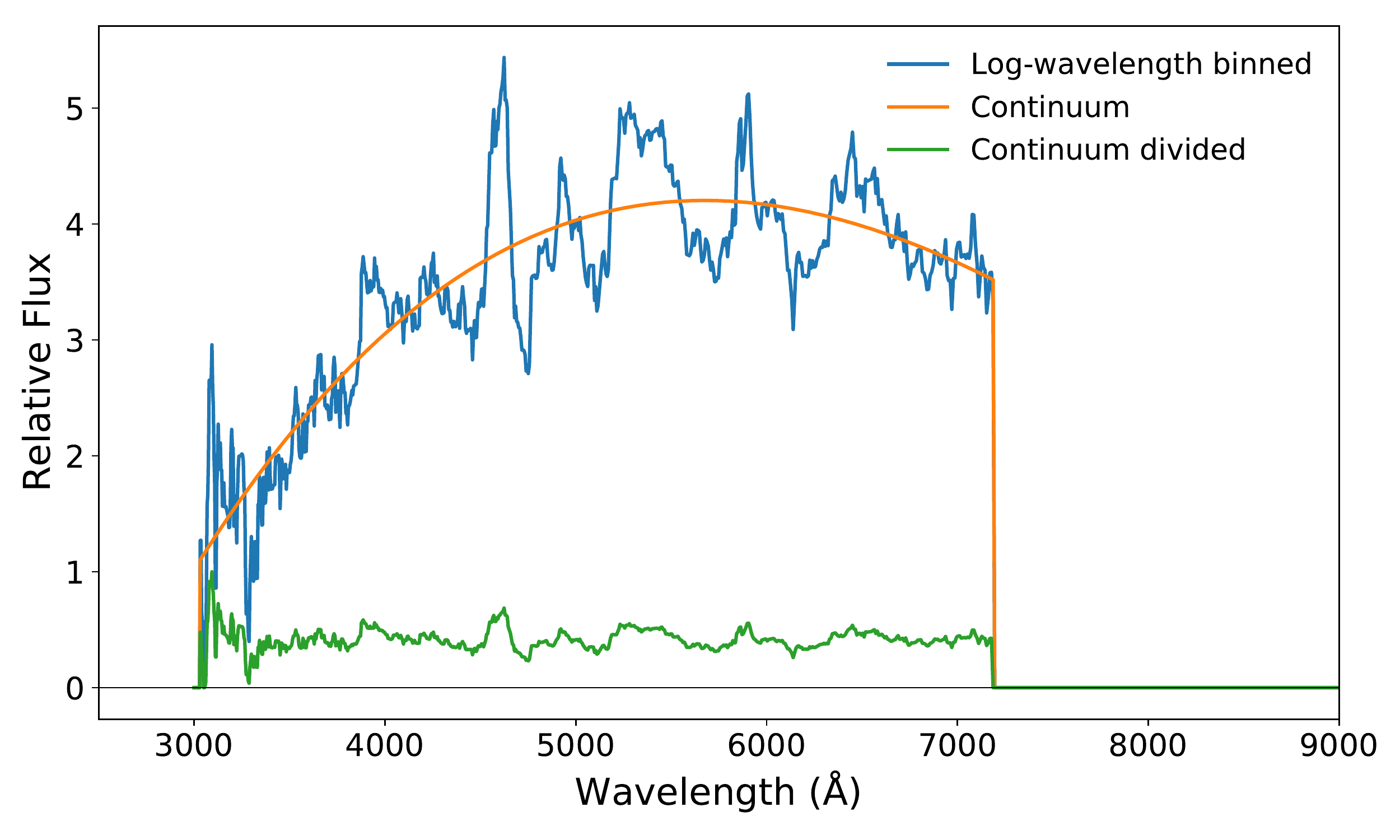}{0.5\textwidth}{(c)}
		\fig{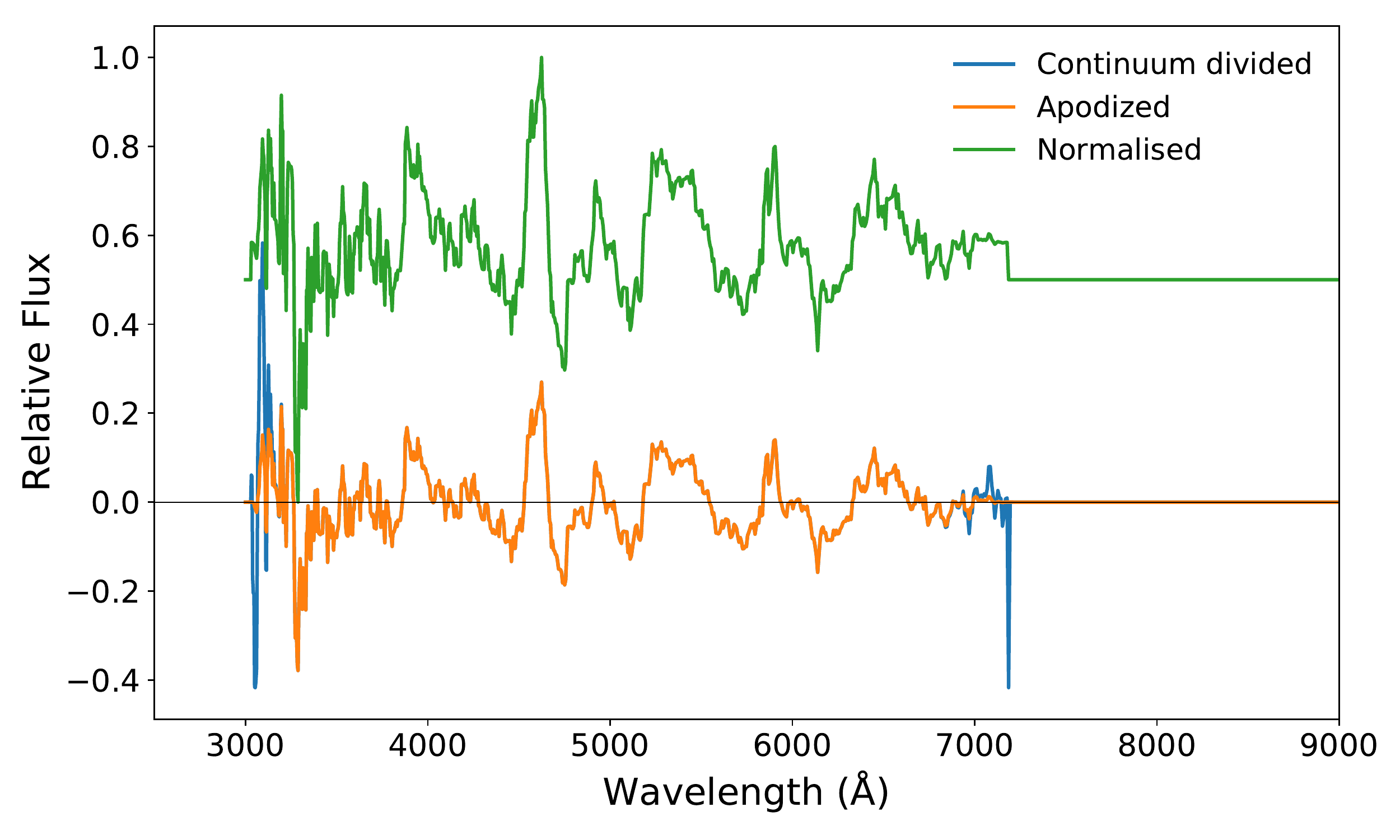}{0.5\textwidth}{(d)}
	}
	\caption{The spectral pre-processing steps before training using the SNIa DES16C2ma spectrum as an example. (a) The blue line is the raw data spectrum, while the orange line shows the result after applying a low-pass median filter with a window size defined by equation \ref{eq:windowSize} and a smoothing factor of 5. (b) The smoothed spectrum is then de-redshifted to its rest frame based on the redshift obtained from its host-lines from an external software. This step is not applied in {\tt DASH} if the redshift-agnostic model is used. It is also binned into $N_w$ points on a log-wavelength scale. (c) The de-redshifted and smoothed spectrum is then binned onto a log-wavelength scale as defined in equation \ref{eq:logwave} (blue line). A 13-point cubic spline interpolation is used to model the continuum (orange line) before it is divided from the binned spectra to remove any spectral colour information (green line). (d) The edge-discontinuities on the previous spectrum is smoothed with a cosine-taper (orange line). The flux is then normalized to values between 0 and 1 (green line).
	\label{fig:preprocessing}}
\end{figure*}

Many of the previous classification and redshifting tools (including {\tt SNID} \citep{Blondin2007}, {\tt MARZ} \citep{Hinton}, and {\tt AUTOZ} \citep{Baldry2014}) pre-process their spectra in a similar way before cross-correlation and template-matching. These methods are loosely based on the algorithms discussed by \citet{Tonry1979}. We implement a very similar processing technique to that used by \citet{Blondin2007} in {\tt SNID}. Our processing algorithm is applied to both the training set and any input spectrum. It consists of the following steps:
\begin{description}
	\item[1. Low pass median filtering] \hfill \\ 
    	The first step to processing is to apply a low-pass median filter to each spectrum in order to remove high-frequency noise and cosmic rays. We scale the amount of smoothing based on the average wavelength spacing of the spectrum, defined as $\lambda_{density}$ below:
        \begin{equation}
        \lambda_{density} = \left( \lambda_{max} - \lambda_{min} \right) / N,
        \end{equation}
        where $\lambda_{min}$ and $\lambda_{max}$ are the minimum and maximum wavelength of the spectrum, and $N$ is the number of points in the spectrum. We also define the wavelength density of the final spectra after processing as:
        \begin{equation}
        w_{density} = \left( w_1- w_0 \right) / N_w.
        \end{equation}
        The window size of the median filter is then defined as:
        \begin{equation}
        \textrm{window\_size} = \frac{w_{density}}{\lambda_{density}} \times  \textrm{smooth},
        \label{eq:windowSize}
        \end{equation}
        where $ \textrm{smooth}$ is the user-defined amount to scale the amount of filtering. Most of the spectra used in the training set have been preprocessed and smoothed by {\tt SNID}, and as such we do not add any further smoothing, and set the window size to 1. Input spectra in {\tt DASH} have a default smoothing factor of $\textrm{smooth} = 6$, but can be altered by a user. An example of  this filtering step is illustrated in Figure \ref{fig:preprocessing}a.
        
	\item[2. De-redshifting] \hfill \\ 
    	The next stage involves de-redshifting the spectrum to its rest frame (illustrated in Figure \ref{fig:preprocessing}b). For input spectra, this is an optional stage depending on which redshift model is used (see section \ref{sec:TrainedModels}).
    	
	\item[3. Log-wavelength binning] \hfill \\ 
    	In the third step, we bin the spectra onto a log-wavelength scale with a fixed number of points ($N_w$) between $w_0$ and $w_1$. These parameters can be changed by a user who wishes to re-train the CNN model. However, the default parameters are $N_w  = 1024$, $w_0 = 3500\AA$, $w_1 = 10000\AA$, which covers the optical spectral range at which most supernova events are observed at, and has enough points to recover both narrow and broad spectral features, while not including too many points to be computationally expensive. These parameters were further selected to match the default parameter values of the {\tt SNID} data, so that we could directly use these in our training set. 
    	
    	This step is important for a few reasons. Firstly, it ensures that each spectrum is a vector of exactly the same length and at the same wavelengths so that vectors from different spectra can be easily compared and trained on. Secondly, it is consistent with the {\tt SNID} data, and can make redshifting less computationally expensive \citep{Blondin2007}. However, perhaps most important, is that we can make use of CNN's natural position invariance \citep{Duda2012} during classification. By using a log-wavelength scale, changes in redshift now become linear translations, and so, the CNN's natural affinity for being invariant to small linear translations can be employed to allow classifications to also be invariant to redshift.

        The log-wavelength binning process follows the same method outlined in \citet{Blondin2007}; some of the key steps are shown here. First, the log-wavelength axis, $w_{\mathrm{log}, n}$, is defined as:
        \begin{equation}
        w_{\mathrm{log}, n} = w_0\ln e ^ {n\times dw_{\mathrm{log}}},
        \label{eq:logwave}
        \end{equation}
        where $n$ is the index of each point in the vector, and runs from $0$ to $N_w$, and
        \begin{equation}
        dw_{\mathrm{log}} = \ln(w_1/w_0)/N_w
        \label{eq:dwlog}
        \end{equation}
        is the size of a logarithmic wavelength bin. The binned wavelength can then be translated from the normal wavelength with the following relationship,
        \begin{equation}
        \textrm{binned\_wave} = A\ln{w_{\mathrm{log}, n}} + B
        \end{equation}
        where $B = -N_w\ln{w_0/\ln(w_1/w_0)}$ and $A = N_w/\ln(w_1/w_0)$.
        Using this method, the input and training spectra were binned onto this scale. The binned spectrum is illustrated as the orange line in Figure \ref{fig:preprocessing}b. \textcolor{black}{The points in the spectrum that do not have data in the range $w_0$ to $w_1$ are set to zero.}

	\item[4. Continuum modelling with spline interpolation] \hfill 
    	The fourth step in preparing the spectra involves dividing the continuum. For galaxy spectra, the continuum is well defined and is easily removed using a least-squares polynomial fit. In supernova spectra, however, the apparent continuum is ill-defined due to the domination of bound-bound transitions in the total opacity \citep{Pinto2000}. For this reason, a 13-point cubic spline interpolation is used to model the continuum. 13 points was considered to be sufficient to interpolate the spectrum. This is illustrated as the orange line on Figure \ref{fig:preprocessing}c.
    	
	\item[5. Continuum division] \hfill \\ 
	    This continuum is then divided from the spectrum (blue line). This step removes any spectral colour information (including flux miscalibrations), and enables the correlation to rely purely on the relative shape and strength of spectral features in each spectrum. It also has the advantage of diminishing the effect of extinction from the remaining spectra. According to \citet{Blondin2007}, the loss of colour information has very little impact on the redshift and age determination.
	    
	\item[6. Apodising the edges] \hfill \\ 
        While the discontinuities at each end of the spectrum are limited by the continuum division, further discontinuities are removed by apodizing the spectrum with a cosine bell in the final step of processing. This involves multiplying 5\% of each end of the spectrum by a cosine, to remove sharp spikes. This is illustrated as the orange line in Figure \ref{fig:preprocessing}d. Finally, the spectrum is renormalised to positive values between 0 and 1 (green line), so that it is ready for training in the CNN. \textcolor{black}{As neural networks require regularly sampled data in a fixed grid, we set the the points in the spectrum that do not have data in the range $w_0$ to $w_1$ to 0.5.}

\end{description}

We then define two important properties for each processed spectrum for the supervised deep learning approach: its label and image data. The image data is composed of the 1024-point vector that corresponds to the pre-processed normalised flux-values. The labels correspond to one of the 306 different classification bins outlined in section \ref{sec:template_distribution}. We represent these labels as 306-point one-hot vectors where each entry represents a different classification bin so that matrix multiplication can be more easily used when training. The labelled and preprocessed data is then passed into the deep learning model for training.

\section{Deep Learning}
\label{sec:DeepLearning}
Deep learning is a branch of machine learning that has recently gained a lot of popularity for its success in a range of different applications including image, speech, and language recognition. The age of big data and advancements in computer hardware have enabled neural networks to be effective at solving these more complicated problems in reasonable amounts of time. 

\subsection{Convolutional Neural Networks}
\label{sec:ConvolutionalNeuralNetworks}
Convolutional neural networks are one of the most popular deep learning architectures, and have been very successful at benchmark image classification problems. We have  employed this architecture by phrasing the spectral classification problem as a one-dimensional image classification problem, where fluxes correspond to pixel intensities. This enables us to use a very similar method to that which is used to solve the benchmark MNIST classification problem \citep{LiDeng2012}. We have developed the CNN with {\tt TensorFlow}'s {\tt Python} library due to its convenient high-level library which avoids low-level details. It makes use of a highly efficient {\tt C++} back-end to do its computations \citep{Abadi2015}.

In a deep neural network, each layer is in the form of a set of nodes or neurons which represent the data. In {\tt DASH}, the first input layer is made up of 1024 neurons representing the fluxes of an input spectrum. Additional layers of neurons above the original input signal are built to ensure that each new layer captures a more abstract representation of the original input layer. Each new hidden layer identifies new features by forming non-linear combinations of the previous layer \citep{Hinton2006,Cybenko1989}. For example, the hidden layers in {\tt DASH} represent abstract constructions of the input flux vector. The final output layer will then simply represent 306 different neurons corresponding to the 306 different classification bins of supernova types and ages. 

The output, $\hat{y}_i$, of each neuron in a  neural network layer can be expressed as the weighted sum of the connections from the previous layer:
\begin{equation}
\hat{y}_i = \sum\limits_{j=1}^{n} W_{i,j} x_j + b_i,
\end{equation}
where $x_j$ are the different inputs to each neuron from the previous layer, $W_{i,j}$ are the weights of the corresponding inputs, $b_i$ is a bias that is added to allow some points in the vector to be more independent of the connections, $j$ is an integer running from 1 to the number of connected neurons in a particular layer to sum over the connections from the previous layer, and $i$ is an integer running from 1 to the number of neurons in the next layer. In the simple case, where we simply have a single layered dense neural network, $x$ is simply the input flux, $i$ runs from 1 to 1024 across the length of the input flux vector, and $j$ runs from 1 to 306 across the number of classification bins. The weights and biases are free variables that are computed by {\tt TensorFlow} during the training process.

In the final output layer, the values of $\hat{y}$ represent the `evidence' tallies for each classification bin. In order to be able to assign probabilities to each of the classification bins, we make use of a softmax regression model in the final layer. The softmax regression probabilities, $y$, are calculated by applying a softmax function on the evidence,
\begin{equation}
y = \mathrm{softmax}(\hat{y}),
\end{equation}
where the softmax activation function is defined as
\begin{equation}
\mathrm{softmax}(x)_i = \frac{e^{x_i}}{\sum\limits _j e^{x_j}}.
\end{equation}

This function generalises a logistic regression to the case where it can handle multiple classes. It effectively normalises the output layer of neurons so that the total probabilities of all classification bins sums to 1. These softmax probabilities are important in {\tt DASH} as they are used to rank the best matching classification bins. It's important to note that these probabilities only provide the relative probability of a particular classification bin when compared to the other 306 different supernova types and ages. 

\begin{figure*}[ht!]
	\includegraphics[width=1\linewidth]{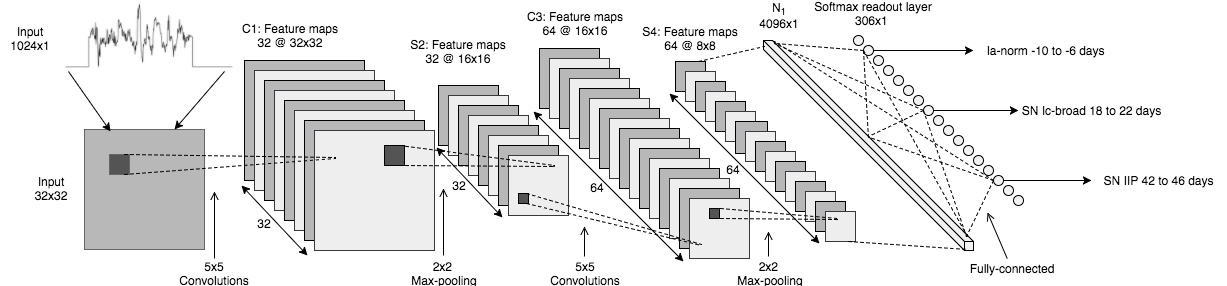}
	\caption{A visual representation of the multilayer convolutional neural network used in {\tt DASH}. The 1024-point input flux, which has been processed following the method outlined in Figure \ref{fig:preprocessing}, is reshaped into a $32\times32$ grid. The first convolutional layer computes 32 features for each $5\times5$ patch on the input. These 32 images are then sub-sampled using a standard max-pooling layer over $2\times2$ patches of each image, reducing the images sizes to $16\times16$. A second layer of convolution with 64 features for each $5\times5$ patch is applied to the previous layer before a $2\times2$ max-pooling layer is used to sub-sample the image size down to $8\times8$. The 64 images representing a $64\times8\times8$ tensor are then flattened down to a 4096-point vector. A fully connected layer with 1024 neurons to allow processing on the entire image is added. Similar to the convolutional layers, weights and biases are computed before a readout and softmax regression layer are added to identify the best matching classifications of the model. This final layer is a 306-point vector, with a score for each supernova type and age bin. Three example classification bins have been listed on the right.}
	\label{fig:deep_learning_model}
\end{figure*}

Before the training process can begin, we need to specify a loss function which indicates how accurately the model's prediction matches the true class for each input spectrum. We define the loss function to be the cross entropy, $H_Y (y)$, between the actual classification bin, $Y$, and the model's prediction, $y$, as:
\begin{equation}
H_{Y} (y) = -\sum\limits_{i=1}^{306} Y_i \log(y_i).
\label{eq:cross-entropy}
\end{equation}
Here, $Y$ is the label of the data which is made up of a 306-point one-hot vector with zeros in all entries except for one which has a 1 to indicate the true classification bin. On the other hand, $y$ is a 306-point vector where the sum of all entries add to 1, and ideally for a good model, the entry with the highest probability would be the same bin as the entry with a 1 in $Y$. Hence, the cross-entropy measures how inefficient the predictions are compared to the truth. We minimise the cross-entropy using a common but sophisticated gradient descent optimiser called the \texttt{Adam} optimiser \citep{KingmaADAMOptimiser}. We feed in our training set defined in section \ref{sec:Data} in small batches and train the neural network such that the values for the weights and biases in each layer are computed to optimise the model.

Overall, the neural network model consists of six different layers: including two convolutional layers with two max-pooling layers between them, one fully connected layer, and a readout layer before the softmax regression as illustrated in Figure \ref{fig:deep_learning_model}. Each convolutional and fully-connected layer have weights and biases which are initialised with a small amount of noise to avoid symmetry-breaking and zero-gradients, and a small positive bias to avoid `dead neurons', respectively. These layers use Rectified Linear Units (ReLU) \citep{Nair2010} as the activation function for each of the neurons in the layers. The max-pooling layers basically just sub-sample the input flux in a non-linear fashion so as to reduce the computational complexity \citep{Boureau2010,Aniyan2017}. Following the fully connected layer, we implement dropout regularisation to reduce over-fitting during training. Effectively, this means that the neurons which have very small weight values, and hence do not strongly interact with other neurons, are discarded from the network iteratively during training.


\section{Trained Models}
\label{sec:TrainedModels}
\subsection{Models}
\label{sec:Models}
Using the machine learning architecture defined in section \ref{sec:DeepLearning}, we have trained four different models which are available in the {\tt DASH} release. All of these models use the same dataset (described in section \ref{sec:Data}) and follow the same data augmentation approaches outlined in section \ref{sec:AugmentationAndOversampling}, whereby various amounts of host galaxy light are added to the data. However, they differ in whether they classify into these hosts, and on whether they calculate the redshift. They are listed as follows:
\begin{enumerate}
	\item Known redshift, SN only classification
	\item Unknown redshift, SN only classification
	\item Known redshift,  SN+host classification
	\item Unknown redshift, SN+host classification
\end{enumerate}

In the majority of supernova spectra, the redshift can be accurately pre-determined from host galaxy features using effective redshifting tools such as {\tt MARZ}\footnote{\url{http://samreay.github.io/Marz/}} \citep{Hinton}. As such, the `Known redshift, SN only' model has been designed with the same CNN architecture illustrated in Figure \ref{fig:deep_learning_model} and ensures that each spectrum in the training set has been de-redshifted to its rest frame ($z=0$). During classification, the redshift must be input as a prior by the user so that the input spectrum can also be de-redshifted to its rest frame.

However, in some supernova spectra, the host galaxy is too faint compared to the supernova spectrum, and hence, the redshift cannot be easily determined from standard redshifting tools. For these cases, we have developed models which can classify supernovae independent of the redshift, and hence do not require a redshift prior. The `Unknown redshift, SN only' model uses the same architecture as the `Known redshift, SN only model', but differs by adding an extra data-augmentation step (see section \ref{sec:AugmentationAndOversampling}) which involves iteratively redshifting each spectrum by varying amounts before training. This enables the trained model to learn the features of spectra independent of their redshift, and hence be able to identify the classification bin regardless of whether the input spectrum is in its rest frame.

Once the best matching classification bins have been identified, we determine the redshift of each of the top ranking classification bins by making use of a cross-correlation technique with the input and the \textcolor{black}{training data} from the classification bin. This redshifting method is described in section \ref{sec:RedshiftingMethods}.

The SN+host classification models are designed with nearly the same architecture as the SN only models, respectively. However, instead of just classifying into the 306 classification bins made up of supernova type and age, we add an extra dimension with 11 host galaxies, making a total of ($11\times306$) 3366 classification bins. In each of these bins we add varying proportions of a particular host galaxy spectrum. The 11 host galaxy \textcolor{black}{types} we used are taken from the {\tt SNID} and BSNIP databases and follow the Hubble diagram naming convention, listed as follows: E, S0, Sa, Sb, Sc, SB1, SB2, SB3, SB4, SB5, SB6. The CNN then trains based on the presence of a combined supernova and host galaxy.

\subsection{Redshifting Methods}
\label{sec:RedshiftingMethods}
\begin{figure}[ht!]
	\includegraphics[width=1\linewidth]{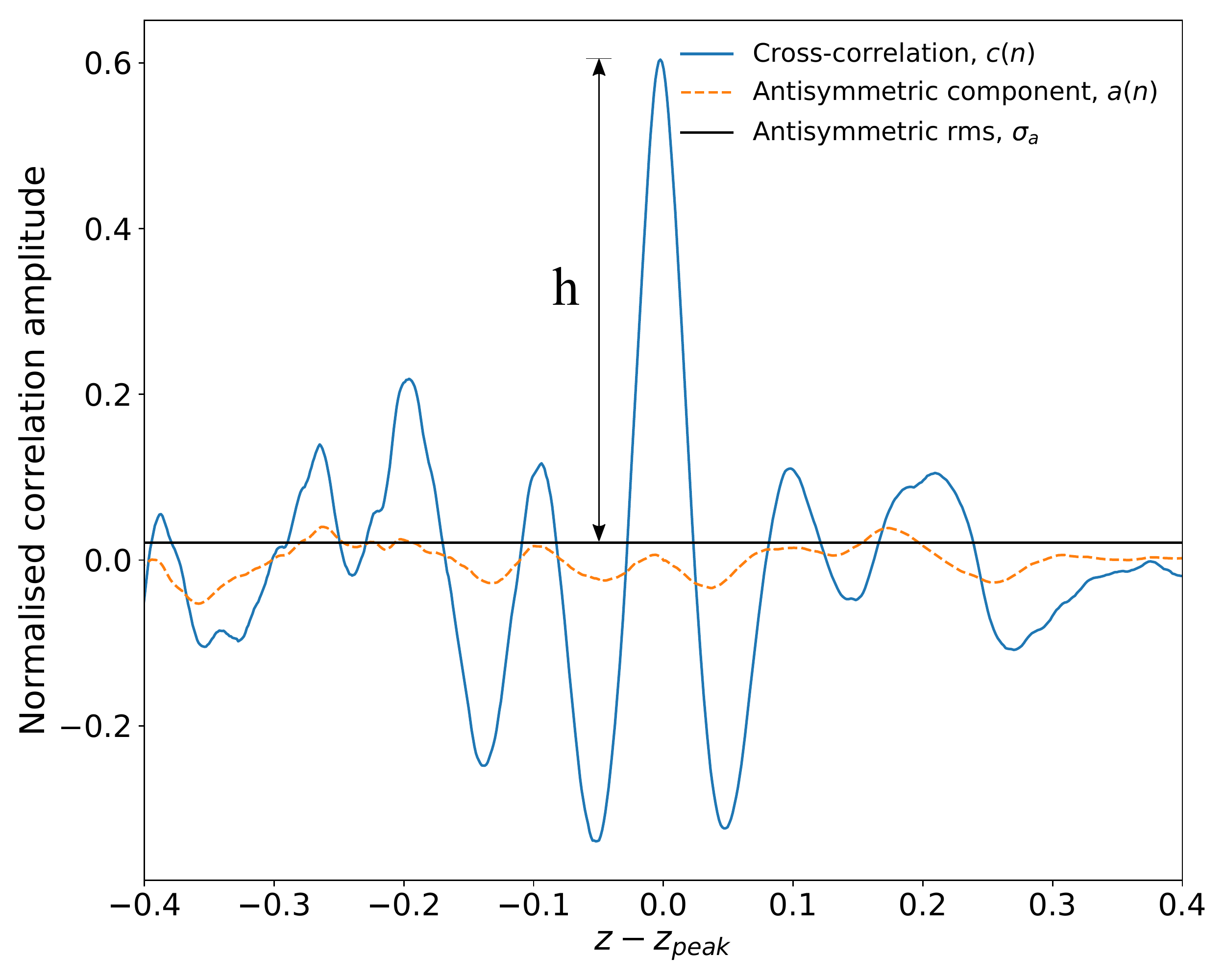}
	\caption{An example cross-correlation function of the DES16C2ma spectrum (Figure \ref{fig:preprocessing}) with the best matching \textcolor{black}{spectrum from the training set} determined by {\tt DASH}. The position of the highest peak is used to determine the redshift with equation \ref{eq:redshift}. The antisymmetric component, $a(n)$, defined in equation \ref{eq:autocorrelation} is shown as the orange dashed line, and the rms of this is illustrated as the horizontal black line. The height of the correlation peak, $h$, is the difference between the highest value and the antisymmetric rms, $\sigma_a$.}
	\label{fig:cross_correlation}
\end{figure}

In the second and fourth models, we iteratively redshift each spectrum in the training set by varying amounts between $z=0$ to $z=1$ before it is trained with the the neural network, hence enabling the model to learn features and classify spectra irrespective of redshift. The log-wavelength scale means that redshifts are now linear translations, and hence, help us to employ the CNN's natural position invariance \citep{Duda2012}. Once the model has determined a best matching classification bin, we calculate the redshift using a very similar cross-correlation technique to that used in {\tt SNID} as defined by \citet{Blondin2007} and \citet{Tonry1979}. The preprocessed input spectrum, $s(n)$, is cross-correlated ($\star$) with each \textcolor{black}{training set spectrum}, $t(n)$, in the classification bin as follows:
\begin{equation}
c(n) = s(n) \star t(n) = \mathcal{F}\left( S(k) T(k) \right),
\end{equation}
where $n$ represents the log wavelength indexes, $c(n)$ is the cross correlation function, $S(k)$ and $T(k)$ represent the fast Fourier transform of the input spectrum and a \textcolor{black}{training set} spectrum, respectively, and $\mathcal{F}$ is the fast Fourier transform function which enables us to calculate the cross-correlation.  An example cross-correlation function of the spectrum used in Figure \ref{fig:preprocessing} and a spectrum \textcolor{black}{from the training set} is illustrated in Figure \ref{fig:cross_correlation}.

Since the spectra have been processed onto a log wavelength scale defined in section \ref{sec:Preprocessing}, the position of the peak cross-correlation score enables the redshift to be computed as
\begin{equation}
z = e^{\delta \times dw_{log}},
\label{eq:redshift}
\end{equation}
where $dw_{log}$ was defined in equation \ref{eq:dwlog} and $\delta$ is the index value of the peak cross-correlation score in the range $-N_w/2 < \delta < N_w/2$. We calculate the redshift from the cross-correlation of the input spectrum with each \textcolor{black}{training set} spectrum in a particular classification bin, and take the median value of all the redshifts. 

The error in the calculated redshift is determined by simply calculating the standard deviation of the redshifts in a particular classification bin.

\subsection{False Positive Rejection}
\label{sec:FalsePositiveRejection}
As outlined in section \ref{sec:ConvolutionalNeuralNetworks}, the ranking system used by {\tt DASH} only provides a relative measure of how closely an input spectrum matches a particular classification bin compared to all other classification bins. If an input spectrum happens to be a weird spectrum, then this ranking system will still choose the closest match, which may lead to false-positive classifications. To account for such cases, we have made use of two independent measures to flag potential misclassifications. The first rejection test makes use of a similar measure to that used in {\tt SNID} called the $rlap$ score, and the second test compares the top ranking {\tt DASH} classifications to ascertain whether the matches are consistent with each other. This provides two independent warnings to a user: a `\textit{low $rlap$ warning}', and an `\textit{inconsistent classification warning}'. These are seen as a `reliability' label in the {\tt DASH} interface, which act to inform a user that the automatic classifications requires closer human inspection.

\subsubsection{Low $rlap$ Warning}
\label{sec:LowRlapWarning}
In {\tt SNID}, the $rlap$ scores act as the primary method of comparing an input spectrum to each of the \textcolor{black}{spectra in the training set}: the \textcolor{black}{training set spectrum} with the highest $rlap$ score is considered the best matching spectrum. 

\citet{Tonry1979} first introduced the cross-correlation height-noise ratio, $r$, to quantify the significance of a cross-correlation peak. It is defined as:
\begin{equation}
r = \frac{h}{\sqrt{2} \sigma_a},
\end{equation}
where $h$ is the height of the cross-correlation shown in Figure \ref{fig:cross_correlation}, and $\sqrt{2}\sigma_a$ is the rms of the antisymmetric component of $c(n)$.  The antisymmetric component is calculated by assuming that $c(n)$ is the sum of an auto-correlation of the \textcolor{black}{training set} spectrum, $t(n)$ with its shifted spectrum $t(n-\delta)$ and a random function, $a(n)$ that distorts the correlation peak \citep{Tonry1979}:
\begin{equation}
c(n) = t(n) \star t(n-\delta) + a(n).
\label{eq:autocorrelation}
\end{equation}
The autocorrelation term will give a peak correlation at the exact redshift given by the shift, $\delta$, in logarithmic wavelength units, and will be symmetrical about $n=\delta$. Assuming that the symmetric and antisymmetric part of $a(n)$ have approximately the same amplitude and are uncorrelated, the rms of $a(n)$ is $\sqrt{2}$ times the rms of its antisymmetric component \citep{Blondin2007}. 

While the $r$ score alone is a sufficient measure of the similarity of two spectra if both the \textcolor{black}{training set spectrum} and input spectra cover a wide wavelength range, it provides a poor measure if the two spectra do not significantly overlap each other in their rest frame. This overlap can be quantified as
\begin{equation}
lap = \ln {\frac{w_a}{w_b}},
\end{equation}
where $w_a$ and $w_b$ are the maximum and minimum wavelengths at which both spectra overlap each other, respectively. Combining the two scores in the product $rlap = r \times lap$ provides a measurement of the similarity between two spectra.

After cross-correlating an input spectrum with each \textcolor{black}{training set spectrum} in the best matching classification bin, we calculate the average value from each of these $rlap$ scores. If the average $rlap$ score is small (defined as $rlap < 6$), then we output a low reliability flag to the user to act as a warning that the automatic classification may not be accurate. This enables a user to more closely inspect the spectra with a low $rlap$ warning. We note that the $rlap$ scores used in {\tt DASH} cannot be directly compared with the scores used in {\tt SNID}.

\subsubsection{Inconsistent Classification Warning}
\label{sec:InconsistentClassifcationWarning}
The second measure of warning a user about a potential misclassification is to compare the top ranking classifications provided by {\tt DASH}. If the top matches are not in neighbouring classification bins, such that the broad supernova type, or the supernova age are distinctly different from each other in the top few matches, then we list the classification with a warning label. More specifically, we check that the top two matches are the same broad supernova type, and also check if the age bins of the top matches are neighbouring each other (i.e. an example of neighbouring bins would be `2 to 6 days' and `6 to 10 days'). If either of these checks fail, then we output a warning to signify that there may be a misclassification.

On the other hand, if the top matching classifications are in agreement, such that they represent the same type of supernova and neighbouring age bins, then we can actually combine the softmax probabilities together to provide a higher level of certainty on the classification. For example, in Figure \ref{fig:DASH_screenshot}, we can combine the top few classifications as they are in neighbouring bins, and output the combined probability, as is illustrated in the top right of the figure.

\section{Performance}
\label{sec:Performance}
In this section we detail the performance of the main Model 1 (see section \ref{sec:Models}) released in {\tt DASH}. The matching algorithms are first validated against the testing set, and then it is tested against recent data taken from 3 years of ATels (Astronomical Telegrams) released by OzDES \citep{2015ATel.8137....1T,2015ATel.8164....1B,2015ATel.8167....1L,2015ATel.8176....1S,2015ATel.8367....1D,2015ATel.8413....1G,2015ATel.8460....1P,2015ATel.8464....1Y,2016ATel.8673....1M,2016ATel.9504....1S,2016ATel.9570....1K,2016ATel.9636....1O,2016ATel.9637....1O,2016ATel.9742....1M,2016ATel.9855....1H,2017ATel.9961....1S,2017ATel10759....1M,2018ATel11146....1C,2018ATel11147....1C,2018ATel11148....1M}.

\subsection{Testing Set}
From the total number of spectra described in section \ref{sec:Data_Description}, initially 80\% was used for training the deep learning algorithm, and 20\% was left for evaluating the matching performance. Once we were confident that the algorithm was effective, we retrained it using 100\% of the data before testing its performance on the OzDES ATels. \textcolor{black}{While it is generally not good practice to apply a model that has not been validated, we decided that in order to be able to classify into classes that are not well represented in the training set, it would be more beneficial to use as much data as we had available. We tested both the validated model (using just 80\% of the data) and the unvalidated model (using all the available data) on the OzDES spectra, and found that while the difference was marginal, the model using all the data produced results that more closely matched the OzDES ATels.}

\begin{figure*}[ht!]
	\plotone{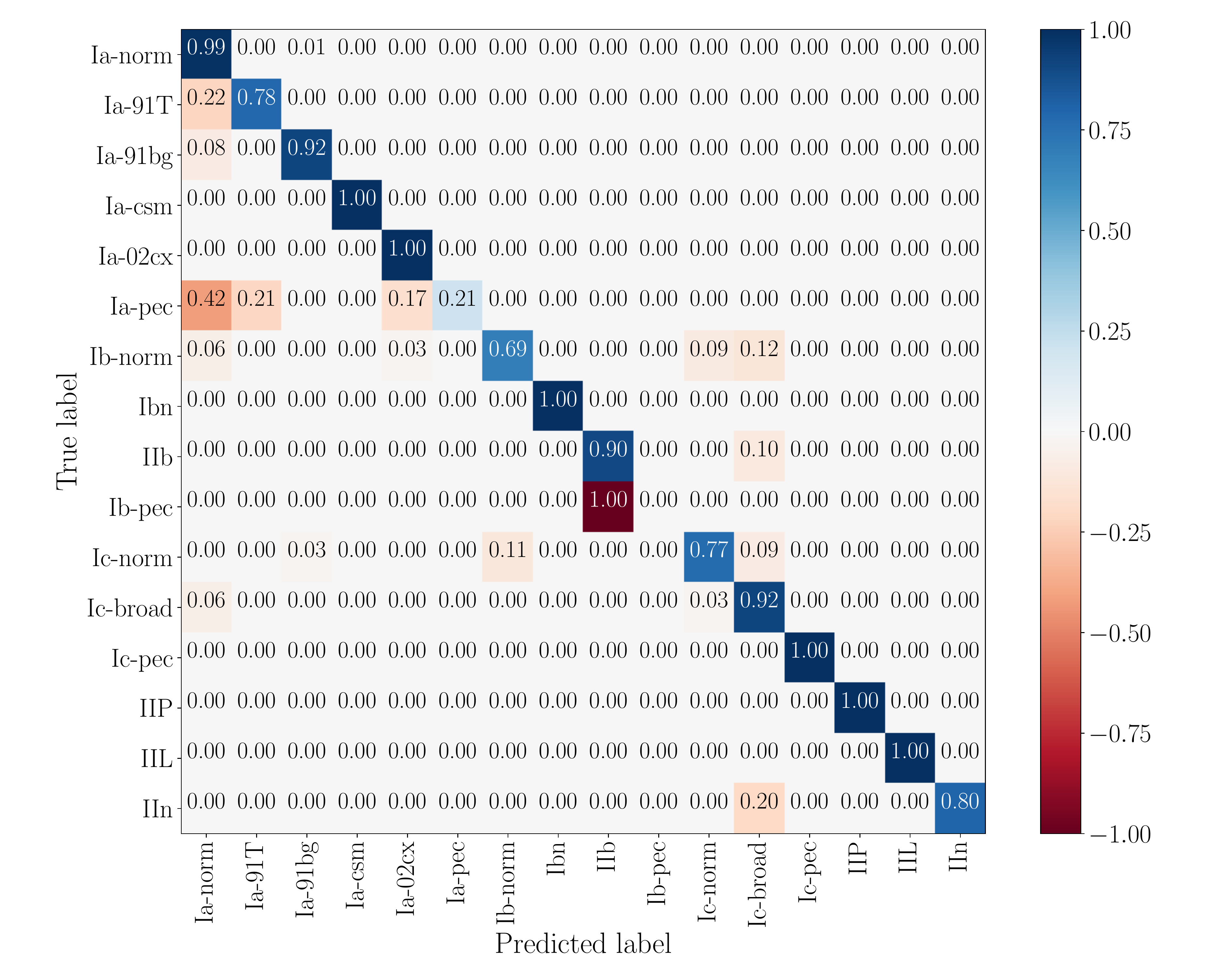}
	\caption{Normalised confusion matrix of the classifier trained on 80\% of the data outlined in section \ref{sec:Data_Description} and tested on the remaining 20\% of spectra. The colour bar and value in each cell indicate the fraction of each \textit{True label} that was classified as the \textit{Predicted Label}. The negative colour bar values indicates misclassifications, while positive corresponds to correct classifications.}
	\label{fig:confusion_matrix}
\end{figure*}

The normalised confusion matrix illustrating the classification performance on the validation set is illustrated in Figure \ref{fig:confusion_matrix}.

The predicted classes are mostly consistent with the true classes, with most misclassifications occurring within the same broad supernova type. For example, the Ia-91T misclassifications were all Ia-norm supernovae and all Ib-pec SNe were misclassified as IIb's. Similarly, there were some SNIb and SNIc misclassifications.

\subsection{OzDES ATels}
To give an indication of how {\tt DASH} will perform on noisy host-contaminated spectra from large surveys based on fibre optics instead of just long-slit spectroscopy, we collected spectra that have been identified in all OzDES ATels from 2015 to 2017 years, and have compared whether {\tt DASH} matches these classifications. This is listed in Table \ref{tab:ATELClassifications} of Appendix \ref{sec:OzDES_atel_comparison}. In the OzDES ATELs, objects are often not classified as precisely as {\tt DASH}, whereby the age of the supernova is not well constrained, and the supernova may be listed with just a broad type without a specific subtype, and may also often be listed with a trailing question mark to indicate that the classifiers were not confident on the classification. Moreover, it should be noted that these classifications were obtained by using data from not just the spectra, but by also making use of the light curve information. To this end, the classifications were completed by two or three experienced astronomers with the help of {\tt Superfit} and {\tt SNID}, but were not autonomously classified like the {\tt DASH} classifications.

{\tt DASH} is able to provide a much more specific classification with the age and subtype constrained with useful probabilities to indicate the confidence of the fit. As a caveat, we note that objects  flagged as \textit{Reliable} should be considered strong classifications even if they have low probabilities because we can sum the probabilities of the next few similar classifications (see section \ref{sec:InconsistentClassifcationWarning}).
 
Furthermore, the speed of classification of {\tt DASH} is significantly better than previous classification tools. Whereby, we were able to autonomously classify all 212 spectra in under 20 seconds, as opposed to the several days to weeks taken to originally classify the objects.

{\tt DASH} was able to classify the entire set of OzDES spectra completely autonomously without any human visual inspection. It matched the ATel classification for 93\% of the spectra, correctly classifying 197 out of the 212 supernovae. \textcolor{black}{These are listed in Table \ref{tab:ATELClassifications} in Appendix \ref{sec:OzDES_atel_comparison}, and summarised in Table \ref{tab:OzDES_summary}. OzDES is primarily a cosmological survey, and thus is biased towards following up SNIa. The OzDES ATels are dominated by type Ia SNe. Only a small fraction of SNII supernovae are included in the ATels (perhaps because these can often be identified by the presence of Hydrogen emission lines), and only three SNIb or SNIc have been included.} 

All but three of the mismatches were either flagged by the False Positive Rejection scheme as \textit{Unreliable} (indicating that the classification should be further checked by a human) or were classified as a `Ic-broad'. In general, we consider that classifications into the Ic-broad class are usually highly host-contaminated spectra, and are not usually actually Ic-broad type SNe. Two of the other three misclassfiications were typed as `SNIa?', indicating that the ATel classifications were uncertain. \textcolor{black}{The accuracy of the classifications coupled with the false positive rejection scheme} ultimately enables astronomers to only need to look at a very small subset of the entire testing set, with most spectra being classified autonomously.

\begin{table}[]
    \centering
    \begin{tabular}{c|c|c}
        \hline
         ATel class & \# of SNe & \# of DASH matches \\ \hline 
         Ia & 129 & 127 \\
         Ia? & 43 & 34 \\
         II & 28 & 25 \\
         II? & 9 & 7 \\
         Ibc & 1 & 1 \\
         Ibc? & 2 & 2 \\
         \hline
    \end{tabular}
    \caption{The distribution of the 212 OzDES ATel classifications released between 2015-2017 is shown in the first two columns. The `?' next to the supernova type was used in the ATel classifications to indicate that the authors of the ATel were not confident in their classification. The third column lists how many of the objects in each class were also correctly classified by DASH. }
    \label{tab:OzDES_summary}
\end{table}

\subsection{Comparison to previous software}
Overall, {\tt DASH} is a more effective classification tool than previous tools for four important reasons: speed, accuracy, classification specificity, and its installation and ease of use.

The main improvement of {\tt DASH} over current tools is its significant speed increase. The primary reason for the increase in speed is that machine learning does not iteratively compare with \textcolor{black}{individual spectra}, but instead classifies based on features in the spectrum. Thus, unlike {\tt SNID} and {\tt Superfit} which increase their computation time linearly with the number of \textcolor{black}{spectra in the training data}, {\tt DASH} is able to separate the training and testing stages. The classification of a single supernova takes only a few seconds in {\tt DASH}, but can take several tens of minutes in {\tt Superfit}. Moreover, while {\tt SNID} is already a fast program, {\tt DASH} is even faster, and this is particularly true when classifying several spectra at once. By making use of the {\tt DASH} library functions, a user is able to classify hundreds or thousands of objects within seconds.

Unlike any other similar software, {\tt DASH} does not iteratively search though and compare an input spectrum to each \textcolor{black}{training set spectrum}. Instead, it learns from the aggregate set of supernovae in a particular classification bin, and trains on the specific features that make up a supernova type using a convolutional neural network. The advantage of this is that a classification is always made based on the entire set of spectra within a particular classification bin, rather than a single spectrum. This reduces the impact of spectra with incorrect classifications or unrepresentative spectra.

Finally, we have made the installation and usage very simple. It can be installed without having to worry about dependencies by making use of the Python Packaging Index. It also enables the simple classification of hundreds of spectra with just two lines of code (see section \ref{sec:PythonLibrary}).

Nonetheless, software like {\tt Superfit} and {\tt SNID} still provide independent classification measures, and used in conjunction with {\tt DASH}, a robust classification scheme can be achieved.

\section{Conclusions}
\label{Conclusions}
We have developed a novel classification tool by using a contemporary convolutional neural network with advanced machine learning techniques. We have diverged from all similar tools which employ either a cross-correlation or chi-squared template matching algorithm. By doing so, we have improved upon previous work to enable {\tt DASH} to be orders of magnitude faster than previous tools, autonomous, more accurate and precise in its classification, and much easier to install and use.

We have collated 4831 supernova spectra from the CfA Supernova Program, BSNIP, and the stripped-envelope collection from Liu and Modjaz. \textcolor{black}{Using this as a training set, we have validated the performance of our classifier on three years of ATels from OzDES. The results indicate that {\tt DASH} is well-suited to classify the large number of spectra soon to be observed by upcoming large scale spectroscopic surveys, such as DESI \citep{DESI} and 4MOST \citep{4MOST}.}

\textcolor{black}{Furthermore, these surveys will have less biased and much more complete samples of supernovae that are better able to capture the diversity in the non-SNIa populations. To improve the classification performance of {\tt DASH} further, we can add this larger and more diverse range of supernovae to our training set. Unlike previous classification tools, increasing the size of the training set does not decrease the classification time. However, the training of the classifier can be computationally expensive, and thus it will be most suitable to retrain {\tt DASH} whenever more spectra that encompass a significantly deeper and wider range of spectral classes becomes available.}

While this is an expansive set, as future surveys increase the supernova catalogue, we can increase the size of our training set to retrain and improve the performance of {\tt DASH} even further. 

A systematic preprocessing algorithm, and data augmentation techniques have enabled us to train a robust learning algorithm. The training of four independent models has further allowed us to classify not only the supernova type and age, but also its host galaxy and redshift.

In a beta version of the {\tt DASH} release, we have included extra Superluminous Supernova (SLSN) classes as a new classification type, and plan to release this in an upcoming version.

Moreover, while we have primarily developed this tool for supernova classification, there is no significant reason why this approach can't be extended to other types of spectra: from different types of stars, galaxies, or AGN in the future.

We have publicly released the software with a graphical interface and a python library available on \texttt{pip} and \texttt{GitHub}, and it has already been used in several published supernova classifications. Ultimately, the speed, accuracy, user-friendliness and versatility of {\tt DASH} presents an advancement to existing spectral classification tools. As such, {\tt DASH} is a viable alternative or complementary spectral classifier for the transient community.

\acknowledgments
DM was supported by an Australian Government Research Training Program (RTP) Scholarship and the Australian Research Council Centre for All-Sky Astrophysics (CAASTRO), through project number CE110001020.
The models were trained with the Obelix supercomputer from the School of Mathematics and Physics at the University of Queensland and the servers at the Research School of Astronomy and Astrophysics at the Australian National University.

\software{{\tt AstroPy} \citep{Robitaille2013},  
	{\tt TensorFlow} \citep{Abadi2015},
	{\tt NumPy} \citep{vanderWalt2011},
	{\tt SciPy} \citep{Jones2001},
	{\tt Qt}.
}

\bibliography{references}

\appendix

\section{Data distribution}
\label{sec:Datadistribution}

\begin{figure}[tbh!]
	\centering
	\includegraphics[width=1\linewidth]{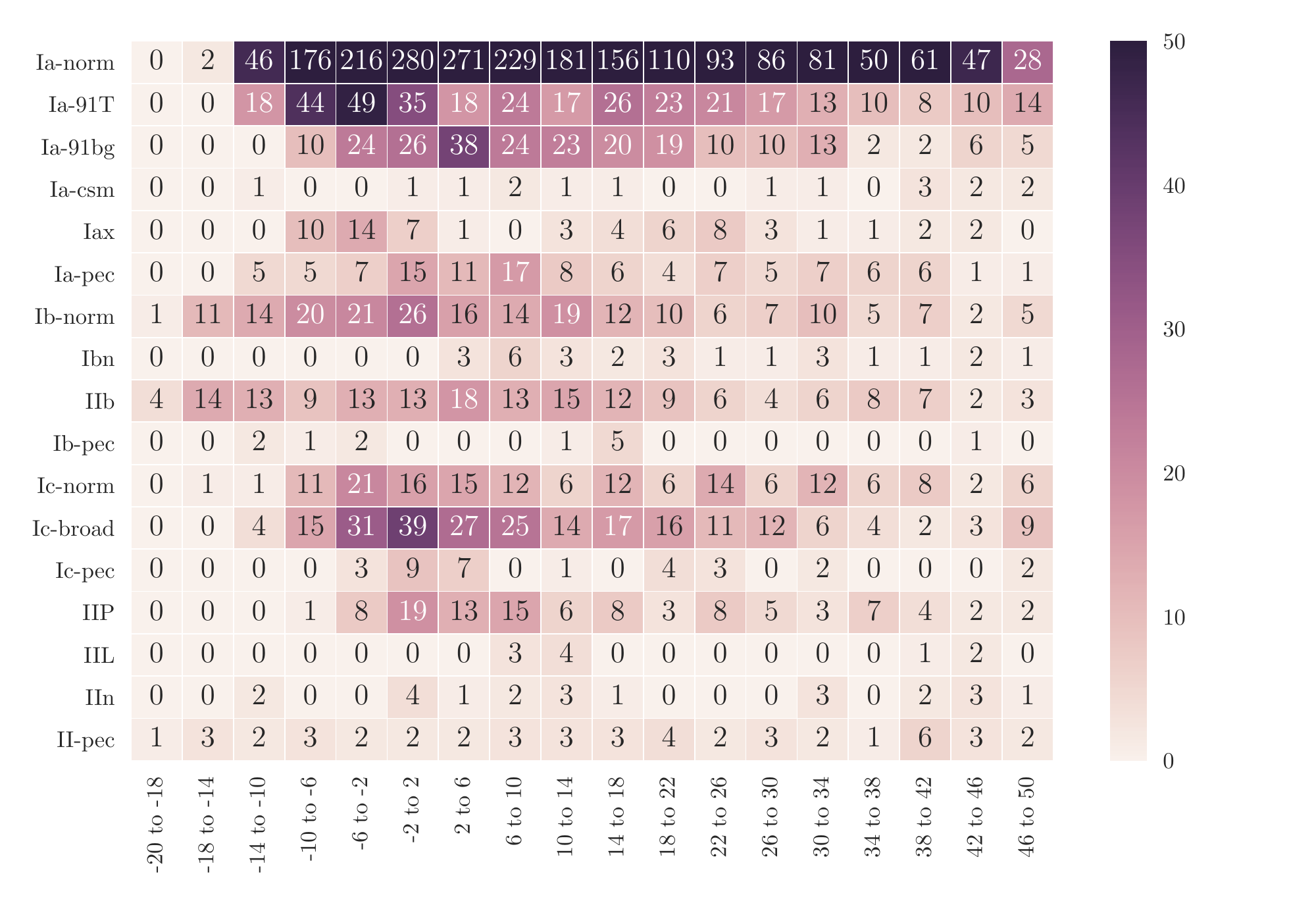}
	\caption{The distribution of the final dataset used to train the machine learning model. The number of spectra for each subtype (rows) and each corresponding age in days since maximum (columns) are listed. The colorbar ranges from 0 to 50 spectra.}
	\label{fig:templatedistribution}
\end{figure}

\section{Usage}
\label{sec:IntendedUse}
{\tt DASH} is intended to be an easy to use supernova classification tool. It has the functionality to quickly classify a single spectrum, but its main advantage over existing tools lies in its ability to automatically classify hundreds or thousands of objects in just a few seconds. As such, it is intended to be used for large scale transient surveys, and is currently being used in the Australian sector of the Dark Energy Survey (OzDES). 

\subsection{Platform}
\label{sec:PlatformAndInterfaces}
We developed {\tt DASH} (Deep Automatic Supernova and Host classifier) as an offline, cross-platform and standalone program based in {\tt Python}. It has been tested to effectively run on most Mac, Linux and Windows distributions, with stringent testing on Mac Sierra and Ubuntu. We have ensured that the installation process is extremely simple, and does not require the messiness of worrying about installing dependencies. The easiest way to install {\tt DASH} is to run \\{\tt pip install astrodash --upgrade} \\ in the command line, which will automatically install nearly every dependency. This simplicity in installation, and the fact that it uses {\tt Python}, which is currently the most popular programming language among astronomers \citep{Momcheva2015}, is a huge advantage compared to previous supernova classification tools.

There are six {\tt Python} based dependencies used in {\tt DASH}, which are all automatically updated and installed with {\tt pip}. We make significant use of Google Brain's new {\tt TensorFlow} {\tt Python} library \citep{Abadi2015} to develop the convolutional neural networks, but also make considerable use of {\tt NumPy} \citep{vanderWalt2011}, {\tt SciPy} \citep{Jones2001}, {\tt AstroPy} \citep{Robitaille2013}, and {\tt Qt} with {\tt PyQt} and {\tt PyQtgraph} for the design of the graphical user interface. The code-base is open source and publicly available on {\tt GitHub}\footnote{\url{https://github.com/daniel-muthukrishna/astrodash/}} and is well-documented\footnote{\url{https://astrodash.readthedocs.io}}.

\subsection{Interfaces}
\label{sec:Interfaces}
Two different interfaces are available in the {\tt DASH} package: a graphical interface and a {\tt Python} library. We detail these in the following subsections.

\subsubsection{Graphical User Interface}

\begin{figure*}[ht!]
	\plotone{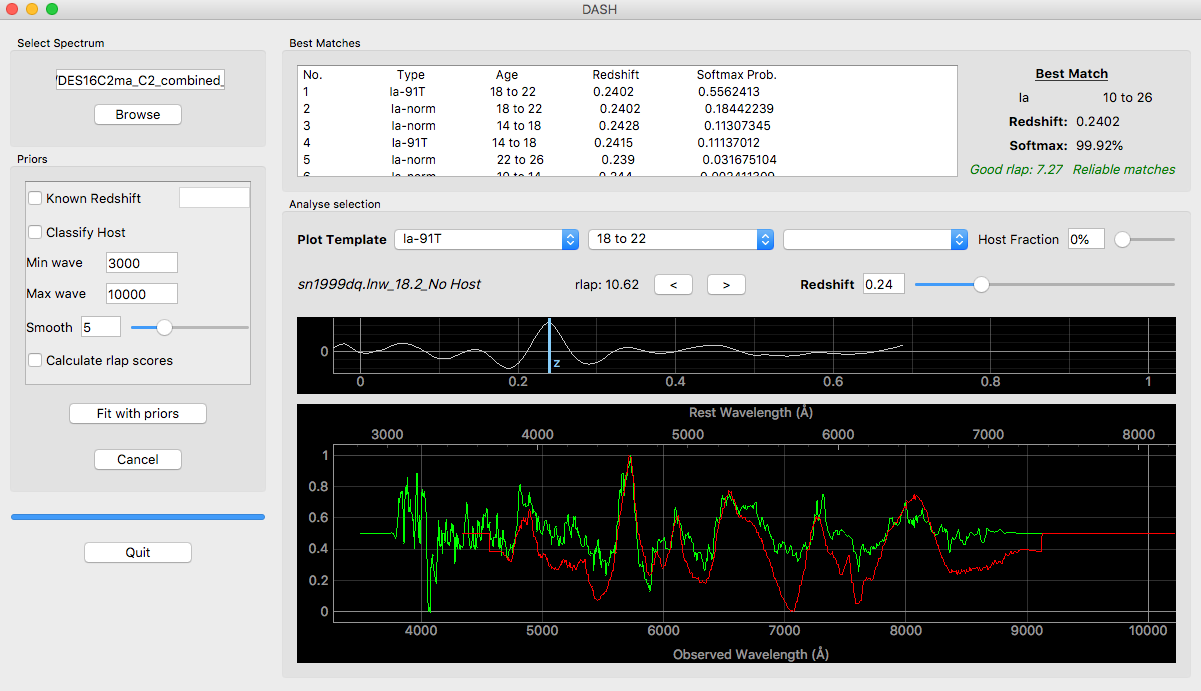}
	\caption{{\tt DASH} graphical interface. An example classification of the OzDES DES16C2ma spectrum (as also illustrated in Figure \ref{fig:preprocessing}) is shown. Using the agnostic redshift model, the software predicts that the input spectrum is a Ia-91T supernova at 18 to 22 days past maximum with a 55.6\% softmax regression confidence. The input spectrum is plotted in the bottom panel (green) against one of the example spectra from the training set (red). The cross correlation is plotted in the smaller graph, with the predicted redshift being $z=0.24$. The probabilities of the top six classifications can be combined, because they are all consistent with each other, to give a combined softmax regression confidence of 99.92\% that the supernova is a SNIa between 10 to 26 days past maximum. Both the rlap and reliable matches flags (see section \ref{sec:FalsePositiveRejection} have passed and are written in green text to indicate that {\tt DASH} is confident about the classification.}
	\label{fig:DASH_screenshot}
\end{figure*}

The graphical interface enables users to visually inspect the {\tt DASH} classifications while being able to tune various parameters. It has been designed to be user-friendly, intuitive, and to contain minimal clutter as illustrated in the example screenshot in Figure \ref{fig:DASH_screenshot}. More detailed instructions on the usage have been provided in the online documentation, but we briefly outline the main components in the following. 

On the left panel under the \textit{Priors} header, the user can make a series of selections which alter the spectrum that is passed into the classification algorithm. The first selection enables the user to choose from one of the four models listed in section \ref{sec:Models} by selecting a combination of the two check boxes. Next, if the user wishes to avoid bad parts of the spectrum caused by excessive noise, dichroic jumps or otherwise, the wavelength range of the input spectrum can be changed. In the case of very noisy spectra, a smoothing option, which applies a low-pass median filter at varying window sizes (as defined in equation \ref{eq:windowSize}), has also been provided. Finally, as well as the softmax probabilities used as a ranking system in {\tt DASH}, users who are familiar with {\tt SNID} may also choose to display $rlap$ values which can act as a second measure of the quality of each classification (see section \ref{sec:LowRlapWarning}). As cross-correlations are relatively slow, checking this box will significantly increase the total classification time. The listed $rlap$ scores are calculated by averaging the scores from the cross correlation of the input spectrum with each \textcolor{black}{training set} spectrum in a particular classification bin.

Once, the priors have been chosen, and the best matching classifications have been filled, the right section of the interface will update to include a few important sections. On the bottom panel, we make use of {\tt PyQtgraph} to plot the preprocessed input spectrum against different \textcolor{black}{training set} spectra. Above this, we also plot the cross-correlation function against redshift for each spectrum similarly to Figure \ref{fig:cross_correlation}. Under the \textit{Best Matches} header, the top ranking classification bins are shown with columns for the type, age, host galaxy, softmax probability, redshift, and $rlap$ score. Depending on the \textit{Priors} selections and the chosen model, only some of these headers will be displayed. On the top right, the best matching classification will be listed by combining the top ranked classifications (as detailed in section \ref{sec:InconsistentClassifcationWarning}). A flag indicating whether the match should be considered reliable or not is also shown based on the false positive rejections tests outlined in section \ref{sec:FalsePositiveRejection}. Under the \textit{Analyse selection} header, a user can choose to plot a different \textcolor{black}{classification bin}, by selecting the type, age, and host of a supernova. Clicking the arrows will switch between the different spectra in a particular classification bin. Finally, the user also has the option to change the fraction of host galaxy light displayed in the \textcolor{black}{training set} spectrum and the redshift of the input to visualise how this affects the spectral features. By default, a spectrum from the best matching classification bin is plotted first.

\subsection{{\tt Python} Library}
\label{sec:PythonLibrary}
A {\tt Python} library has also been developed so that several classifications can be made autonomously without the requirement of visual inspection. Classification of multiple spectra is very simple, requiring only a couple of lines of code:
\begin{lstlisting}
import astrodash

classify = astrodash.Classify(filenames, redshifts)
print(classify.list_best_matches())
astrodash.plot_with_gui(indexToPlot=0)
\end{lstlisting}
The only inputs required are a list of filenames containing the spectra which are to be classified, and an optional list of corresponding known redshifts. More optional arguments selecting which model should be applied, the amount of smoothing, and whether rlap scores should be calculated can also be specified. The details of these optional arguments are outlined in the documentation. However, it should be noted that with just those two lines of code, several hundreds of spectra can be classified automatically and within just a few seconds or minutes, with the best matches being saved to a human-readable text file. The final line enables the first spectral file in the input list to be plotted and analysed on the graphical interface.

\subsection{Usage with Open Supernova Catalogs}
\label{sec:OSC}
{\tt DASH} also interfaces with the online Open Supernova Catalog\footnote{\url{https://sne.space/}} \citep{OSC}. Changing the filename input in either interface with something in the format osc-name-age\_index (e.g. osc-sn2002er-10) will download the spectrum from the OSC, and classify it.

\subsection{Development and Contribution}
\label{sec:DevelopmentAndContribution}
The {\tt DASH} source code currently consists of several thousand lines of code across more than 30 {\tt Python} files which are open-source and publicly hosted on a {\tt git} repository on {\tt GitHub} at \url{https://github.com/daniel-muthukrishna/astrodash}.

{\tt GitHub} provides issue tracking to keep track of open issues and feature requests. Users are encouraged to report bugs or issues, and to request new useful features with this issue tracker. Moreover, this project has been developed in an object oriented fashion, so that different code implementations can be relatively easily changed. One such example is the ability to easily change the deep learning architecture by just replacing one {\tt Python} file. To this end, as more advanced neural network architectures become available, the learning algorithm can be improved or replaced. 

Furthermore, as more supernova spectra are observed by large scale surveys, the training set should be updated. In fact, the more spectra that we can train the CNN with, the better the classification algorithm will become. To this end, if any users of the software would like to increase the size of the training set, they should contact us so that better models can be trained. Alternatively, simply updating the spectra in the training\_set directory on {\tt GitHub} and carefully running the `create\_and\_save\_data\_files.py' file will begin to train a new model. It should be noted, that this training process may take a significant amount of computation time: usually on the order of hours depending on the computational resources available.

Finally, at the time of writing, the project has just the lead author as the sole active developer of the software. However, if users of the software would like to implement their own features which may be useful to others, we encourage them to contact us so that we can add them to the {\tt GitHub} collaborators.

\section{OzDES ATel classification comparison}
\label{sec:OzDES_atel_comparison}
\LTcapwidth=\textwidth
\begin{longtable*}{c|c|c|c|c|c|c}
\hline
 \multirow{2}{*}{\textbf{Name}} & \multirow{2}{*}{\textbf{Redshift}} & \multirow{2}{*}{\textbf{ATel}} & \multicolumn{3}{c}{\textbf{DASH}} & \multirow{2}{*}{\textbf{Match?}} \\ 
& & \textbf{Classification} & \textbf{Classification} & \textbf{Probability} & \textbf{Reliability} \\
\hline
 DES15X3hp & 0.236 &  Ia  +3 weeks & Ia-norm (18 to 22) & 0.835 & Reliable & \checkmark\\
 DES15X3dyu & 0.425 &  Ia   max  & Ia-norm (-10 to -6) & 0.938 & Reliable & \checkmark\\
 DES15X3auw & 0.151 &  Ia  +2 weeks & Ia-norm (10 to 14) & 0.994 & Reliable & \checkmark\\
 DES15X1bw & 0.13 &  Ia  +5 weeks & Ia-91T (42 to 46) & 0.984 & Reliable & \checkmark\\
 DES15E2nk & 0.308 &  Ia  +1 week  & Ia-91T (6 to 10) & 0.947 & Reliable & \checkmark\\
 DES15E2atw & 0.147 &  Ia  +1 week  & Ia-norm (18 to 22) & 0.911 & Reliable & \checkmark\\
 DES15C3fx & 0.2 &  Ia  +3 weeks & Ia-csm (10 to 14) & 0.999 & Reliable & \checkmark\\
 DES15S2dye & 0.26 &  Ia  +1 week  & Ia-norm (-2 to 2) & 0.953 & Reliable & \checkmark\\
 DES15S1by & 0.129 &  II  post-max & Ic-broad (-6 to -2) & 0.999 & Unreliable & x \\
 DES15C3edd & 0.36 &  Ia   max  & Ia-91T (-2 to 2) & 0.707 & Reliable & \checkmark\\
 DES15C2dyj & 0.395 &  Ia  +1 week  & Ia-norm (2 to 6) & 0.955 & Reliable & \checkmark \\
 DES15C2eaz & 0.062 &  II   max  & IIP (2 to 6) & 0.905 & Reliable & \checkmark\\
 DES15C2aty & 0.149 &  Ia  +2 weeks & Ia-norm (18 to 22) & 0.405 & Reliable & \checkmark\\
 DES15C1atm & 0.207 &  Ia  +2 weeks & Ia-norm (18 to 22) & 0.59 & Reliable & \checkmark\\
 DES15X3kqv & 0.142 &   Ia    at max  & Ia-norm (2 to 6) & 0.996 & Reliable & \checkmark\\
 DES15E1kwg & 0.105 &   Ia    at max  & Ia-norm (-6 to -2) & 0.709 & Reliable & \checkmark\\
 DES15X1ith & 0.16 &   Ia    +1 week  & Ia-norm (6 to 10) & 0.905 & Reliable & \checkmark\\
 DES15E1kvp & 0.442 &   Ia    At max             & Ia-norm (2 to 6) & 1.0 & Reliable & \checkmark\\
 DES15C3efn & 0.077 &   Ia    +2 weeks  & Ia-norm (14 to 18) & 0.964 & Reliable & \checkmark\\
 DES15X3iv & 0.018 &   Ia    +1 month  & Ia-norm (46 to 50) & 1.0 & Reliable & \checkmark\\
 DES15X3itc & 0.338 &   Ia    +2-5 days post max & Ia-91T (-2 to 2) & 1.0 & Reliable & \checkmark\\
 DES15X3kxu & 0.345 &   Ia    At max             & Ia-norm (-2 to 2) & 0.94 & Reliable & \checkmark\\
 DES15E1iuh & 0.105 &   II    at max   & IIP (6 to 10) & 0.918 & Reliable & \checkmark\\
 DES15X2asq & 0.28 &   Ia   +7 days  & Ia-91T (6 to 10) & 0.655 & Reliable & \checkmark\\
 DES15S2ar & 0.247 &   Ia?  +16 days & Ia-norm (10 to 14) & 0.977 & Reliable & \checkmark\\
 DES15C1eat & 0.45 &   Ia?  +7 days  & Ia-norm (2 to 6) & 0.612 & Reliable & \checkmark\\
 DES15X1ebs & 0.58 &   Ia?  pre-max  & Ic-broad (2 to 6) & 0.605 & Reliable & x \\
 DES15C3bj & 0.287 &   II?  post-max & Ic-broad (-6 to -2) & 0.807 & Reliable & \checkmark\\
 DES15C3axd & 0.42 &   Ia?    max    & Ia-pec (2 to 6) & 0.997 & Reliable & \checkmark\\
 DES15C1ebn & 0.41 &   Ia?    max    & Ic-broad (2 to 6) & 0.904 & Reliable & \checkmark\\
 DES15C3lvt & 0.4 &   Ia?  post-max & Ic-broad (2 to 6) & 0.596 & Unreliable & x \\
 DES15E2cwm & 0.291 &   Ia?  +10 days & Ia-91T (10 to 14) & 0.964 & Reliable & \checkmark\\
 DES15S2og & 0.38 &   Ia?  post-max & Ia-norm (6 to 10) & 0.744 & Reliable & \checkmark\\
 DES15S1cj & 0.166 &   II?  post-max & IIP (6 to 10) & 0.832 & Reliable & \checkmark\\
 DES15S1ebd & 0.408 &   Ia?    max    & Ia-91bg (2 to 6) & 0.504 & Unreliable & \checkmark\\
 DES15S2dyb & 0.56 &   Ia?  +7 days  & Ic-broad (-10 to -6) & 0.514 & Reliable & x \\
 DES15X3flq & 0.368 &   Ia?  +10 days & Ia-91bg (-2 to 2) & 0.5952 & Unreliable & \checkmark\\
 DES15E2kvn & 0.208 &   Ia?    max    & Ia-norm (-2 to 2) & 0.971 & Reliable & \checkmark\\
 DES15C2lpp & 0.181 &   II?  post-max & Ic-broad (2 to 6) & 1.0 & Unreliable & x \\
 DES15C2lna & 0.069 &   II   post-max & IIn (10 to 14) & 0.897 & Unreliable & \checkmark\\
 DES15X2lxw & 0.197 &   Ia    +1 week & Ia-norm (2 to 6) & 0.984 & Reliable & \checkmark\\
 DES15X2mei & 0.248 &   Ia     max    & Ia-norm (-6 to -2) & 0.527 & Reliable & \checkmark\\
 DES15X3lya & 0.29 &   Ia     max    & Ia-norm (2 to 6) & 0.709 & Reliable & \checkmark\\
 DES15S1mjm & 0.26 &   Ia?  +1 week  & Ia-norm (2 to 6) & 0.574 & Reliable & \checkmark\\
 DES15S1lyi & 0.359 &   Ia?  +3 weeks & Ia-pec (10 to 14) & 0.602 & Reliable & \checkmark\\
 DES15S2mpl & 0.257 &   Ia   +1 week  & Ia-norm (2 to 6) & 0.998 & Reliable & \checkmark\\
 DES15S2mpg & 0.186 &   Ia     max    & Ia-91T (2 to 6) & 0.402 & Reliable & \checkmark\\
 DES15E1neh & 0.39 &   Ia?     max   & Ic-norm (2 to 6) & 1.0 & Unreliable & x \\
 DES15X3mpq & 0.188 &   II   +1 month & IIP (10 to 14) & 0.832 & Unreliable & \checkmark\\
 DES15X2mpm & 0.235 &   Ia   +2 week  & Ia-pec (10 to 14) & 0.964 & Reliable & \checkmark\\
 DES15X2mku & 0.09 &   II   +1 month & IIP (6 to 10) & 0.999 & Reliable & \checkmark\\
 DES15C1mvy & 0.32 &   Ia      max   & Ia-91T (6 to 10) & 0.98 & Reliable & \checkmark\\
 DES15C1mqf & 0.111 &   Ia   +3 weeks & Ia-norm (14 to 18) & 0.959 & Reliable & \checkmark\\
 DES15X3naa & 0.331 &   Ia   -4 days  & Ia-91T (6 to 10) & 0.724 & Reliable & \checkmark\\
 DES15X3nad & 0.1 &   II      max   & IIP (2 to 6) & 0.508 & Unreliable & \checkmark\\
 DES15X2mzv & 0.313 &   Ia      max   & Ia-91T (6 to 10) & 0.929 & Reliable & \checkmark\\
 DES15X2nkl & 0.304 &   Ia      max   & Ia-norm (-2 to 2) & 0.797 & Reliable & \checkmark\\
 DES15C2njv & 0.181 &   Ia      max   & Ia-norm (2 to 6) & 0.559 & Reliable & \checkmark\\
 DES15C1nfb & 0.13 &   Ia      max   & Ia-norm (2 to 6) & 0.57 & Reliable & \checkmark\\
 DES15E2nlz & 0.41 &   Ia   -5 days  & Ia-norm (-2 to 2) & 0.789 & Reliable & \checkmark\\
 DES15E1nei & 0.313 &   Ia      max   & Ia-norm (2 to 6) & 0.97 & Reliable & \checkmark\\
 DES15C3mpk & 0.182 &   Ia   +10 days & Ia-91T (6 to 10) & 1.0 & Reliable & \checkmark\\
 DES15C2oxo & 0.336 &   Ia?  +9 days  & Ia-norm (6 to 10) & 0.69 & Reliable & \checkmark\\
 DES15C3orz & 0.18 &   Ia   +5 days  & Ia-91bg (10 to 14) & 0.844 & Reliable & \checkmark\\
 DES15C3olc & 0.067 &   Ia   +24 days & Ia-norm (18 to 22) & 1.0 & Reliable & \checkmark\\
 DES16X1ey & 0.076 &  SNII   post-max & IIn (6 to 10) & 0.818 & Unreliable & \checkmark\\
 DES16C3ea & 0.217 &  SNIa   post-max & Ia-norm (14 to 18) & 0.925 & Unreliable & \checkmark\\
 DES16E1ah & 0.149 &  SNII   post-max & Ia-norm (26 to 30) & 0.999 & Reliable & x \\
 DES16E1md & 0.178 &  SNIa     max    & Ia-91T (-2 to 2) & 0.982 & Reliable & \checkmark\\
 DES16C3bq & 0.241 &  SNIa     max    & Ia-norm (-2 to 2) & 1.0 & Reliable & \checkmark\\
 DES16C3fv & 0.322 &  SNIa   -6 days  & Ia-norm (-10 to -6) & 0.546 & Reliable & \checkmark\\
 DES16X3jj & 0.238 &  SNII?  post-max & IIL (10 to 14) & 1.0 & Unreliable & \checkmark\\
 DES16X3es & 0.554 &  SNIa?    max    & Ia-pec (2 to 6) & 0.996 & Unreliable & \checkmark\\
 DES16X3hj & 0.308 &  SNIa     max    & Ia-91T (-2 to 2) & 0.899 & Reliable & \checkmark\\
 DES16X3er & 0.167 &  SNIa   +2 days  & Ia-91T (-2 to 2) & 1.0 & Reliable & \checkmark\\
 DES16X3km & 0.054 &  SNII   post-max & IIP (6 to 10) & 1.0 & Reliable & \checkmark\\
 DES16E2dd & 0.075 &  SNIa   +3 days  & Ia-norm (2 to 6) & 0.964 & Reliable & \checkmark\\
 DES16E1de & 0.292 &  SNIa?  +2 days  & Ia-norm (-10 to -6) & 0.568 & Unreliable & \checkmark\\
 DES16X2auj & 0.144 &   Ia   max      & Ia-norm (2 to 6) & 0.94 & Reliable & \checkmark\\
 DES16X1ge & 0.25 &   Ia   post-max & Ia-norm (22 to 26) & 0.988 & Reliable & \checkmark\\
 DES16C2ma & 0.24 &   Ia   post-max & Ia-norm (18 to 22) & 0.658 & Reliable & \checkmark\\
 DES16C2aiy & 0.182 &   Ia   post-max & Ia-norm (2 to 6) & 0.993 & Reliable & \checkmark\\
 DES16X3biz & 0.24 &   Ia   pre-max  & Ia-norm (-6 to -2) & 0.982 & Reliable & \checkmark\\
 DES16X3aqd & 0.033 &   II-P post-max & IIb (-14 to -10) & 0.961 & Reliable & \checkmark\\
 DES16E2aoh & 0.403 &   Ia   post-max & Ia-91T (-6 to -2) & 0.864 & Reliable & \checkmark\\
 DES16C3bq & 0.237 &   Ia   post-max & Ia-norm (2 to 6) & 0.868 & Reliable & \checkmark\\
 DES16E2bht & 0.392 &  SNIa   +3 days  & Ia-norm (-2 to 2) & 0.988 & Reliable & \checkmark\\
 DES16E2bkg & 0.478 &  SNIa     max    & Ia-norm (-2 to 2) & 0.599 & Reliable & \checkmark\\
 DES16X2crt & 0.57 &  SNIa?  near-max  & Ia-norm (-2 to 2) & 0.622 & Unreliable & \checkmark\\
 DES16E2cjg & 0.48 &  SNIa   near-max  & Ia-91T (-6 to -2) & 0.563 & Reliable & \checkmark\\
 DES16X3cpl & 0.205 &  SNII?  near-max  & IIn (-2 to 2) & 0.991 & Unreliable & \checkmark\\
 DES16C3at & 0.217 &  SNII   +60 days  & Ic-broad (2 to 6) & 0.881 & Reliable & x \\
 DES16C1bnt & 0.351 &  SNIa   +1 month  & Ia-norm (22 to 26) & 0.994 & Reliable & \checkmark\\
 DES16C2cbv & 0.109 &  SNII   near-max  & IIP (2 to 6) & 0.822 & Reliable & \checkmark\\
 DES16C1cbg & 0.111 &  SNII   post-max  & IIP (-2 to 2) & 0.999 & Reliable & \checkmark\\
 DES16X2bvf & 0.135 &  SNIb   post-max  & Ib-norm (14 to 18) & 0.65 & Reliable & \checkmark\\
 DES16X2cpn & 0.28 &  SNIa   +1 week   & Ia-norm (6 to 10) & 1.0 & Reliable & \checkmark\\
 DES16X2crr & 0.312 &  SNIa   near-max  & Ia-norm (-2 to 2) & 0.996 & Reliable & \checkmark\\
 DES16X2bkr & 0.159 &  SNII   post-max  & IIP (22 to 26) & 0.953 & Reliable & \checkmark\\
 DES16X2ceg & 0.335 &  SNIa   near-max  & Ia-norm (6 to 10) & 0.709 & Unreliable & \checkmark\\
 DES16E2cqq & 0.426 &  SNIa   -1 week   & Ia-norm (-2 to 2) & 0.362 & Reliable & \checkmark\\
 DES16E2clk & 0.367 &  SNIa   near-max  & Ia-91T (-2 to 2) & 0.99 & Reliable & \checkmark\\
 DES16E2crb & 0.229 &  SNIa   near-max  & Ia-norm (2 to 6) & 0.998 & Reliable & \checkmark\\
 DES16S1cps & 0.274 &  SNIa   -1 week   & Ia-91T (-6 to -2) & 0.718 & Reliable & \checkmark\\
 DES16E1ciy & 0.174 &  SNIa   near-max  & Ia-norm (2 to 6) & 0.992 & Reliable & \checkmark\\
 DES16X2dqz & 0.204 & SNIb/c?   max    & Ic-norm (-2 to 2) & 0.993 & Reliable & \checkmark\\
 DES16X1der & 0.453 &  SNIa   +1 week  & Ia-norm (2 to 6) & 0.982 & Reliable & \checkmark\\
 DES16S2drt & 0.331 &  SNIa     max    & Ia-91T (-10 to -6) & 0.737 & Reliable & \checkmark\\
 DES16E1eef & 0.32 &  SNIa     max    & Ia-91T (-2 to 2) & 0.582 & Reliable & \checkmark\\
 DES16E1eae & 0.534 &  SNIa     max    & Ia-norm (-2 to 2) & 0.994 & Unreliable & \checkmark\\
 DES16X1dbx & 0.345 &  SNIa   +1 week  & Ia-91T (6 to 10) & 0.942 & Reliable & \checkmark\\
 DES16S2dfm & 0.3 &  SNIa   near-max & Ia-norm (2 to 6) & 1.0 & Reliable & \checkmark\\
 DES16S2ean & 0.161 &  SNIa   pre-max  & Ia-norm (-6 to -2) & 0.772 & Reliable & \checkmark\\
 DES16X1dbw & 0.336 &  SNIa   +1 week  & Ia-91T (6 to 10) & 1.0 & Reliable & \checkmark\\
 DES16X1drk & 0.463 &  SNIa   near-max & Ia-norm (-2 to 2) & 1.0 & Reliable & \checkmark\\
 DES16E2drd & 0.27 &  SNIa   near-max & Ia-norm (-2 to 2) & 0.927 & Reliable & \checkmark\\
 DES16E2cxw & 0.293 &  SNIa   +2 weeks & Ia-norm (10 to 14) & 0.826 & Reliable & \checkmark\\
 DES16C3dhv & 0.3 &  SNIa   near-max & Ia-norm (2 to 6) & 0.999 & Reliable & \checkmark\\
 DES16X3dfk & 0.15 &  SNIa   near-max & Ia-norm (2 to 6) & 0.986 & Reliable & \checkmark\\
 DES16E1dic & 0.207 &  SNIa     max    & Ia-norm (2 to 6) & 0.985 & Reliable & \checkmark\\
 DES16E1dcx & 0.453 &  SNIa   +2 weeks & Ia-norm (10 to 14) & 0.928 & Reliable & \checkmark\\
 DES16S2ffk & 0.373 &  SNIa?   -1 week   & Ia-91T (-2 to 2) & 0.754 & Reliable & \checkmark\\
 DES16X1chc & 0.043 &  SNIa    +2 months & Ic-norm (34 to 38) & 0.965 & Reliable & \checkmark\\
 DES16X1few & 0.311 &  SNIa    -1 week   & Ia-norm (-6 to -2) & 0.946 & Reliable & \checkmark\\
 DES16X2dzz & 0.325 &  SNIa?   +2 weeks  & Ia-norm (10 to 14) & 0.771 & Reliable & \checkmark\\
 DES16C1fgm & 0.361 &  SNIa    -4 days   & Ia-91T (-2 to 2) & 0.979 & Reliable & \checkmark\\
 DES16S1ffb & 0.164 &  SNIa    near-max  & Ia-norm (-6 to -2) & 0.552 & Reliable & \checkmark\\
 DES16X3enk & 0.331 &  SNIa?   +1 week   & Ia-norm (10 to 14) & 0.725 & Reliable & \checkmark\\
 DES16X3eww & 0.445 &  SNIa?     max     & Ia-norm (-2 to 2) & 0.995 & Reliable & \checkmark\\
 DES16C2ege & 0.348 &  SNIa?   +1 month  & Ic-norm (10 to 14) & 0.999 & Reliable & x \\
 DES16X3dvb & 0.329 &  SNII    near-max  & Ic-broad (-10 to -6) & 0.895 & Unreliable & \checkmark\\
 DES16C3elb & 0.429 &  SNIa    +1 week   & Ic-norm (10 to 14) & 0.666 & Unreliable & x \\
 DES17E2ci & 0.127 &  SNII    post-max  & IIn (42 to 46) & 0.764 & Reliable & \checkmark\\
 DES17E2ce & 0.269 &  SNIa    +3 weeks  & Ia-norm (18 to 22) & 0.969 & Reliable & \checkmark\\
 DES17E1by & 0.287 &  SNIa    -1 week   & Ia-norm (-2 to 2) & 0.7 & Reliable & \checkmark\\
 DES17E2bx & 0.272 &  SNIa     at-max   & Ia-91T (-2 to 2) & 0.999 & Reliable & \checkmark\\
 DES17E2bw & 0.147 &  SNIa    +2 weeks  & Ia-norm (10 to 14) & 0.998 & Reliable & \checkmark\\
 DES17E2ar & 0.513 &  SNIa    +10 days  & Ia-csm (6 to 10) & 0.944 & Unreliable & \checkmark\\
 DES17E2aq & 0.352 &  SNIa    -1 week   & Ia-norm (-10 to -6) & 0.848 & Reliable & \checkmark\\
 DES17E2b & 0.227 &  SNIa    +1 month  & Ia-norm (18 to 22) & 0.518 & Reliable & \checkmark\\
 DES17E2a & 0.295 &  SNIa?   +1 month  & Ia-norm (22 to 26) & 0.989 & Reliable & \checkmark\\
 DES17X3ct & 0.206 &  SNIbc?  post-max  & Ib-norm (-6 to -2) & 0.983 & Unreliable & \checkmark\\
 DES17X3cb & 0.317 &  SNIa     at-max   & Ia-norm (-2 to 2) & 0.981 & Reliable & \checkmark\\
 DES17X3ca & 0.198 &  SNIa?   +6 weeks  & Ia-norm (38 to 42) & 0.585 & Unreliable & \checkmark\\
 DES17X3bd & 0.141 &  SNII?   post-max  & IIP (26 to 30) & 0.986 & Reliable & \checkmark\\
 DES17X3az & 0.56 &  SNIa?   +1 week   & Ia-91T (6 to 10) & 0.397 & Unreliable & \checkmark\\
 DES17C3eg & 0.117 &  SNIa    +3 weeks  & Ia-norm (22 to 26) & 0.977 & Reliable & \checkmark\\
 DES17C3de & 0.107 &  SNII    post-max  & IIP (38 to 42) & 0.656 & Reliable & \checkmark\\
 DES17E2cc & 0.149 &  SNII    post-max  & IIP (18 to 22) & 0.985 & Reliable & \checkmark\\
 DES17S2byx & 0.31 &  SNIa?   pre-max   & Ia-norm (-10 to -6) & 0.997 & Reliable & \checkmark\\
 DES17E2bhj & 0.186 &  SNII?   post-max  & Ib-norm (-6 to -2) & 1.0 & Unreliable & x \\
 DES17X3bhi & 0.39 &  SNIa    -1 week   & Ia-norm (-6 to -2) & 0.922 & Reliable & \checkmark\\
 DES17X3btv & 0.407 &  SNIa    near-max  & Ia-91T (-6 to -2) & 0.51 & Reliable & \checkmark\\
 DES17S2als & 0.388 &  SNIa    +1 month  & Ia-norm (18 to 22) & 0.885 & Unreliable & \checkmark\\
 DES17E1byv & 0.378 &  SNIa    pre-max   & Ia-norm (-2 to 2) & 1.0 & Reliable & \checkmark\\
 DES17E2bro & 0.223 &  SNIa    -1 week   & Ia-91T (-2 to 2) & 0.554 & Reliable & \checkmark\\
 DES17X1boi & 0.565 &  SNIa    near-max  & Ib-norm (-6 to -2) & 0.999 & Unreliable & \checkmark\\
 DES17E1bmf & 0.566 &  SNIa    near-max  & Ic-broad (-10 to -6) & 0.98 & Reliable & x \\
 DES17C3biz & 0.23 &  SNIa    pre-max   & Ia-norm (-10 to -6) & 0.94 & Reliable & \checkmark\\
 DES17E1axa & 0.237 &  SNIa    +2 weeks  & Ia-norm (14 to 18) & 0.931 & Reliable & \checkmark\\
 DES17X1ayb & 0.292 &  SNIa    +2 weeks  & Ia-91T (6 to 10) & 0.702 & Reliable & \checkmark\\
 DES17X1axb & 0.139 &  SNII    +10 days  & IIP (18 to 22) & 0.947 & Reliable & \checkmark\\
 DES17X1aow & 0.139 &  SNII    post-max  & IIP (18 to 22) & 0.99 & Reliable & \checkmark\\
 DES17X1alj & 0.24 &  SNIa?   +3 weeks  & Ia-norm (22 to 26) & 0.946 & Unreliable & \checkmark\\
 DES17E2sp & 0.312 &  SNIa    +1 month  & Ia-norm (18 to 22) & 1.0 & Reliable & \checkmark\\
 DES17C3dw & 0.17 &  SNII    post-max  & IIP (6 to 10) & 1.0 & Reliable & \checkmark\\
 DES17X1gd & 0.189 &  SNII?   post-max  & IIP (6 to 10) & 0.651 & Unreliable & \checkmark\\
 DES17S2bph & 0.362 &  SNIa?   near-max  & Ia-91T (-2 to 2) & 0.981 & Reliable & \checkmark\\
 DES17S2bop & 0.385 &  SNIa    near-max  & Ia-norm (2 to 6) & 0.859 & Reliable & \checkmark\\
 DES17E2boo & 0.288 &  SNIa    near-max  & Ia-csm (6 to 10) & 0.858 & Reliable & \checkmark\\
 DES17X2bmp & 0.466 &  SNIa?   +1 week   & Ia-norm (2 to 6) & 0.801 & Reliable & \checkmark\\
 DES17E2bmb & 0.44 &  SNIa    near-max  & Ia-norm (-2 to 2) & 0.992 & Reliable & \checkmark\\
 DES17X2blx & 0.344 &  SNIa      max     & Ia-norm (2 to 6) & 0.64 & Reliable & \checkmark\\
 DES17C1azd & 0.338 &  SNIa      max     & Ia-csm (6 to 10) & 0.737 & Reliable & \checkmark\\
 DES17X2bfi & 0.34 &  SNIa    pre-max   & Ia-norm (-2 to 2) & 0.801 & Reliable & \checkmark\\
 DES17E2arn & 0.38 &  SNIa    +2 weeks  & Ia-pec (2 to 6) & 1.0 & Reliable & \checkmark\\
 DES17X2alq & 0.38 &  SNIa?   +2 weeks  & Ia-norm (18 to 22) & 0.997 & Unreliable & \checkmark\\
 DES17X2agh & 0.306 &  SNIa?   +2 weeks  & Ia-norm (18 to 22) & 0.946 & Reliable & \checkmark\\
 DES17S2oo & 0.23 &  SNII    post-max  & IIP (34 to 38) & 0.861 & Unreliable & \checkmark\\
 DES17S2lg & 0.339 &  SNIa?   +1 month  & Ia-norm (22 to 26) & 0.621 & Unreliable & \checkmark\\
 DES17X2abj & 0.252 &  SNII?   post-max  & IIP (2 to 6) & 0.984 & Reliable & \checkmark\\
 DES17E1bud & 0.552 &  SNIa?   near-max  & Ia-norm (-10 to -6) & 0.998 & Unreliable & \checkmark\\
 DES17C2bqz & 0.61 &  SNIa?   near-max  & Ia-norm (-6 to -2) & 0.971 & Unreliable & \checkmark\\
 DES17E1bqq & 0.463 &  SNIa      max     & Ia-norm (-2 to 2) & 0.558 & Reliable & \checkmark\\
 DES17S1bof & 0.226 &  SNIa    pre-max   & Ia-91T (-6 to -2) & 1.0 & Reliable & \checkmark\\
 DES17E1bis & 0.251 &  SNIa    +1 week   & Ia-91T (6 to 10) & 0.996 & Reliable & \checkmark\\
 DES17E1beg & 0.222 &  SNIa    +1 week   & Ia-norm (6 to 10) & 0.915 & Reliable & \checkmark\\
 DES17S1aya & 0.306 &  SNIa?   pre-max   & Ia-csm (-14 to -10) & 0.985 & Reliable & \checkmark\\
 DES17S1bch & 0.136 &  SNIa      max     & Ia-norm (2 to 6) & 0.995 & Reliable & \checkmark\\
 DES17C2acb & 0.35 &  SNIa?   +2 weeks  & Ia-norm (18 to 22) & 0.898 & Reliable & \checkmark\\
 DES17C2pf & 0.135 &  SNII    post-max  & II-pec (38 to 42) & 1.0 & Reliable & \checkmark\\
 DES17C2ou & 0.103 &  SNIa   +2 months  & Ia-norm (46 to 50) & 0.991 & Unreliable & \checkmark\\
 DES17S1lu & 0.084 &  SNII    post-max  & IIP (38 to 42) & 0.998 & Reliable & \checkmark\\
 DES17C1bql & 0.195 &  SNIa    -1 week   & Ia-norm (-6 to -2) & 0.761 & Reliable & \checkmark\\
 DES17C3blq & 0.511 &  SNIa?     max     & Ic-broad (2 to 6) & 0.344 & Unreliable & x \\
 DES17C3bei & 0.103 &  SNII    near-max  & IIb (-2 to 2) & 0.665 & Reliable & \checkmark\\
 DES17C1bat & 0.197 &  SNIa    +2 weeks  & Ia-norm (6 to 10) & 1.0 & Reliable & \checkmark\\
 DES17C3aye & 0.157 &  SNII    post-max  & IIb (-18 to -14) & 0.977 & Reliable & \checkmark\\
 DES17C1ayc & 0.435 &  SNIa    +2 weeks  & Ia-91bg (-2 to 2) & 0.93 & Unreliable & \checkmark \\
 DES17C1ald & 0.131 &  SNIa    post-max  & Ia-norm (22 to 26) & 0.822 & Reliable & \checkmark\\
 DES17S1emx & 0.185 &  SNIa?    -1 week   & Ia-91T (-10 to -6) & 0.549 & Reliable & \checkmark\\
 DES17S2ebs & 0.304 &  SNIa     at max    & Ia-norm (-2 to 2) & 0.535 & Reliable & \checkmark\\
 DES17C3dxw & 0.622 &  SNIa?    near-max  & Ic-norm (2 to 6) & 1.0 & Reliable & x \\
 DES17X1dyt & 0.33 &  SNIa     -1 week   & Ia-91T (-6 to -2) & 0.992 & Reliable & \checkmark\\
 DES17X2dwm & 0.3 &  SNIa     near-max  & Ia-norm (2 to 6) & 1.0 & Reliable & \checkmark\\
 DES17X1dwi & 0.252 &  SNIa     at max    & Ia-91T (-2 to 2) & 0.593 & Reliable & \checkmark\\
 DES17X1diq & 0.625 &  SNIa?    near-max  & Ib-norm (-6 to -2) & 0.744 & Unreliable & x \\
 DES17X1cuy & 0.55 &  SNIa?    +1 week   & Ib-norm (-6 to -2) & 0.842 & Unreliable & x \\
 DES17X3dub & 0.123 &  SNII     near-max  & IIP (22 to 26) & 0.696 & Unreliable & \checkmark\\
 DES17E1dgn & 0.453 &  SNIa     near-max  & Ia-norm (-6 to -2) & 0.96 & Reliable & \checkmark\\
 DES17C3doq & 0.32 &  SNIa     at max    & Ia-91T (-2 to 2) & 0.909 & Reliable & \checkmark\\
 DES17C1cpv & 0.19 &  SNIa     +1 week   & Ia-norm (6 to 10) & 0.992 & Reliable & \checkmark\\

\hline 
\caption{Classification of supernovae released in the past 3 years of ATels by OzDES \citep{2015ATel.8137....1T,2015ATel.8164....1B,2015ATel.8167....1L,2015ATel.8176....1S,2015ATel.8367....1D,2015ATel.8413....1G,2015ATel.8460....1P,2015ATel.8464....1Y,2016ATel.8673....1M,2016ATel.9504....1S,2016ATel.9570....1K,2016ATel.9636....1O,2016ATel.9637....1O,2016ATel.9742....1M,2016ATel.9855....1H,2017ATel.9961....1S,2017ATel10759....1M,2018ATel11146....1C,2018ATel11147....1C,2018ATel11148....1M}. The first column is the name of the observed object, the second column is the redshift determined by {\tt MARZ}. The third column is the classification given in the ATel by OzDES. It details the type and age from maximum. A question mark after the classification type indicates that the ATel was not certain on the classification. Most of these ATel classifications were made by the OzDES team with the help of {\tt Superfit} or {\tt SNID}. The fourth, fifth and sixth columns are the classification, softmax regression probability and reliability from {\tt DASH}, respectively. The final column has a tick if the ATel and {\tt DASH} agree on the type of the supernova, and a cross if they disagree.}
\label{tab:ATELClassifications}
\end{longtable*}

\end{document}